\documentclass{article}
\usepackage{xcolor}
\usepackage{ulem}
\usepackage{amsmath}
\usepackage{graphicx}
\usepackage{enumerate}
\usepackage{natbib}
\usepackage{url}
\usepackage{tikz}
\usepackage{subfigure}
\usepackage{multirow}
\usepackage{amssymb}
\usepackage{enumitem}
\usepackage{tablefootnote}

\newcommand{\bfbeta}{\mbox{\boldmath $\beta$}}

\newcommand{\bfs}{\mbox{$\mathbf{s}$}}
\newcommand{\bfalpha}{\mbox{\boldmath $\alpha$}}
\newcommand{\bfeta}{\mbox{\boldmath $\eta$}}
\newcommand{\bfnu}{\mbox{\boldmath $\nu$}}

\newcommand{\bfSigma}{\mbox{\boldmath $\Sigma$}}
\newcommand{\bfgamma}{\mbox{\boldmath $\gamma$}}

\graphicspath{{./FiguresTB/}}

\begin{document}

	\def\spacingset#1{\renewcommand{\baselinestretch}%
		{#1}\small\normalsize} \spacingset{1}


	\title{\bf The impact of directly observed therapy on the efficacy of Tuberculosis treatment: A Bayesian multilevel approach}
	\author{Widemberg S. Nobre\thanks{Department of Statistical Methods, Federal University of Rio de Janeiro, Rio de Janeiro - RJ, Brazil.}, Alexandra M. Schmidt\thanks{Department of Epidemiology, Biostatistics and Occupational Health, McGill University, Montreal - QC, Canada.}, Erica E. M. Moodie\footnotemark[2] \hspace{.1cm} and \\ David A. Stephens\thanks{ Department of Mathematics and Statistics, McGill University, Montreal - QC, Canada.}\\
		\date{}
	}
	\maketitle
	
	\bigskip
	
	\begin{abstract}
We propose and discuss a Bayesian procedure to estimate the average treatment effect (ATE) for multilevel observations in the presence of confounding.  We focus on situations where the confounders may be latent (e.g., spatial latent effects). This work is motivated by an interest in determining the causal impact of directly observed therapy (DOT) on the successful treatment of Tuberculosis (TB); the available data correspond to individual-level information observed across different cities in a state in Brazil.
We focus on propensity score regression and covariate adjustment to balance the treatment (DOT) allocation. We discuss the need to include latent local-level random effects in the propensity score model to reduce bias in the estimation of the ATE. A simulation study suggests that accounting for the multilevel nature of the data with latent structures in both the outcome and propensity score models has the potential to reduce bias in the estimation of causal effects.
	\end{abstract}
	
	\noindent%
	\vfill
	
	\spacingset{1.25}
	
\section{Motivation}\label{sec:motivation}

Tuberculosis (TB) is an infectious disease transmitted through air and caused by {\it Mycobacterium tuberculosis} bacteria. Transmission occurs when an uninfected person inhales airborne particles containing {\it Mycobacterium tuberculosis} bacteria, which become airborne when a person with laryngeal or pulmonary TB coughs, speaks, or sneezes, for example. There are effective treatments against TB; in Brazil, antibiotics are provided by the Brazilian public health system ({\it Sistema \'Unico de Sa\'ude}) for at least six months.

Despite the availability of proven treatments, TB remains a leading cause of death. In 2018 alone, TB was responsible for the death of approximately 1.5 million people globally. One of the most problematic issues in the treatment of TB is drug resistance acquired by the bacteria in patients, which is usually due to the mismanagement of medications (\url{https://www.who.int/features/qa/79/en/}).  In the early 1990's, the World Health Organization (WHO) introduced \textit{directly observed therapy} (DOT) in an effort to improve the probability of success of the treatment of TB. DOT relies on a relationship between a patient and a health professional wherein the professional observes or assists the patient ingesting the medications. In Brazil, DOT is characterized as patients being assisted at least three times per week during treatment. The objectives of DOT are to reduce emotional consequences and to educate patients about TB \citep{reis2015directly,reis2019tuberculosis}.

The efficacy of  DOT in the treatment of TB has been debated (see e.g., \cite{weis1994effect,chaulk1998directly,lienhardt2012global}). \cite{anuwatnonthakate2008directly} found, using propensity score methods, that starting the treatment with DOT tends to reduce the odds of treatment failure.
Their methods did not consider the potential for unmeasured confounding, i.e., they assumed that treatment assignment is strongly ignorable \citep{rosenbaum1983central}. 
However, to investigate the causal impact of DOT on the efficacy of treatment, the assumption of strong ignorability may not be reasonable. \cite{monteiro2017use} note that in the state of S\~ao Paulo, socio-economic factors impact both the seeking of and access to health assistance. Their results suggest that reports of chronic diseases (including TB) are more common in people with lower socio-economic status as compared to those with higher socio-economic status. They observe that the predominant reasons reported for no access to health care services included long waiting periods and lack of health professionals. Furthermore, \cite{paim2011brazilian} suggest that the public health sector in Brazil is influenced by political issues related to decentralization and dependence on support systems. The issues discussed in \cite{monteiro2017use} and \cite{paim2011brazilian} are closely linked to the risk factors for drug resistance noted in  \cite{henrique2020determinants} and suggest the potential for unmeasured confounding.

Our interest in this paper is a quantification of the effect of DOT on rates of TB treatment success. The available data were gathered from different municipalities (cities) across the state of S\~ao Paulo, Brazil, and comprises information for 12,057 individuals. {The dataset includes individuals who followed treatment for TB and had a concluding diagnosis recorded in 2016 by the Brazilian public health system}. These data were made available by the Brazilian public health system. 

\begin{figure}[!htb]
	\centering
	\includegraphics[scale=.58]{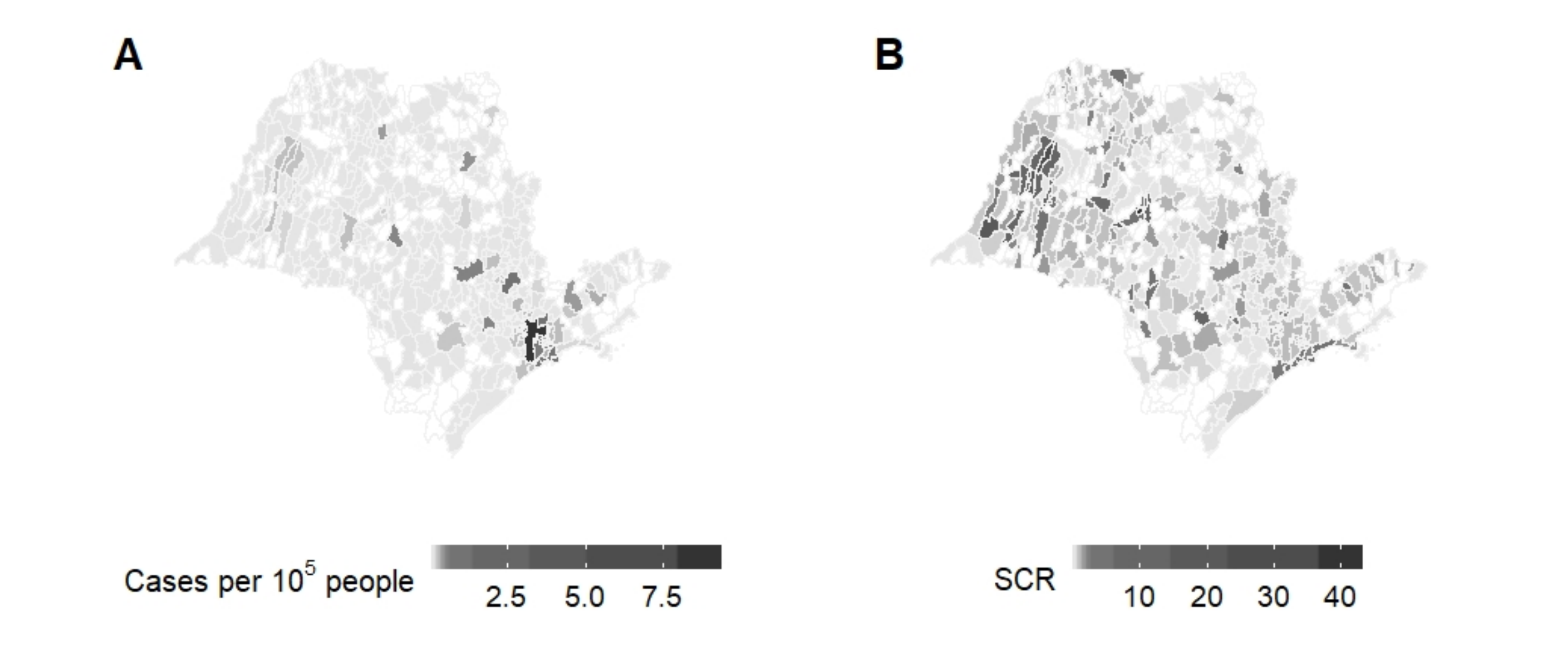}
	\caption{The state of S\~ao Paulo with cities colored by the incidence of TB cases per 100,000 people (Panel A) and the SCR (Panel B). Cities in white are those that did not have any cases of TB with a concluding diagnosis in the dataset.}
	\label{fig:SPmap}
\end{figure}

The state of S\~ao Paulo is divided into 645 municipalities and has approximately 46 million inhabitants, of whom nearly 40 million live in urban areas. The state capital, S\~ao Paulo, is the most populous city, with approximately 12 million inhabitants and a demographic density of $7.398,26\; {\rm hab/km}^2$ (\url{https://cidades.ibge.gov.br/brasil/sp/sao-paulo/panorama}). The observed cases of TB that had a concluding diagnosis in 2016 are spread over 415 municipalities, and 30\% of the cases were observed in the city of S\~ao Paulo. The maps in Figure \ref{fig:SPmap} depict the incidence of TB cases per 100,000 people and the standardized cases ratio (SCR) across the state of S\~ao Paulo. 
The SCR was evaluated based on cities with at least one case of TB with a concluding diagnosis in 2016 present in the data; regions in white recorded no cases of TB with a concluding diagnosis in 2016. The SCR for a particular city is defined as the number of cases divided by the number of expected cases for each city. From Panel B, the western region of S\~ao Paulo state shows the highest SCR of TB in the state.

Our goal is to estimate the average treatment effect (ATE) of DOT on the rate of cure for TB under a Bayesian framework while accounting for unmeasured confounding that may vary across the cities of S\~ao Paulo state. As individual-level data are available for each municipality and DOT is administered via the public health system, {it is expected that individuals residing in the same municipality share common aspects of the TB treatment; such as those suggested by \cite{monteiro2017use} and \cite{paim2011brazilian} which were noted above.} This raises the question of whether municipal-level random effects should be included when modeling the quantities of interest. Therefore, we follow the Bayesian paradigm and discuss the use of multilevel approaches to propensity score and outcome models.

In the causal framework, observational studies are characterized by a joint distribution for the triple ($\mathbf{X}$,$Z$,$Y$) that can distort the causal relationships between the exposure, $Z$, and the outcome, $Y$, due to confounding.
Confounding results from stochastic relationships among $\mathbf{X}$, $Z$, and $Y$, that arise due to the mechanism that generates the observed data. In particular, data generation involves the set of covariates, $\mathbf{X}$, that affect both the treatment assignment and the outcome. The focus of our analyses is on the ATE, which is defined as
\begin{equation}
	\begin{split}
		\tau &= E\left\{Y(1) - Y(0)\right\}= E_{\mathbf{X}}\{E_{Y|Z,\mathbf{X}}(Y|Z=1,\mathbf{X}) - E_{Y|Z,\mathbf{X}}(Y|Z=0,\mathbf{X})\},
	\end{split}
	\label{eq:ateTB}
\end{equation}
where $Y({\rm z})$ denotes the potential outcome that would be observed if we intervene to set $\{Z = {\rm z}\}$ \citep{rubin1974estimating}.
{ Consistent estimation of $\tau$ follows under a correct (parametric) specification of $E_{Y|Z,\mathbf{X}}(Y|Z={\rm z},\mathbf{X})$.} If correct specification cannot be assured, estimates of $\tau$ might be biased due to confounding. Further, integration over $\mathbf{X}$ is required to represent a population-level estimand. In practice, this integration is usually achieved by empirically averaging over the observed data. { Such causal estimation presumes that no acquisition (also known as selection) bias is present in the data.}  

A natural approach to overcome confounding is to use regression with a balancing score \citep{rubin1974estimating}. A balancing score is defined as any function $B(\mathbf{X})$ such that $Z$ and $\mathbf{X}$ are conditionally independent given  $B(\mathbf{X})$. The propensity score is one such balancing score for binary exposures. Assuming that there is no unmeasured confounding, the ATE can be computed through the inclusion of the propensity score in a regression model of $Y$ given $Z$ \citep{rosenbaum1983central}. 

Assuming that strong ignorability holds (i.e., the potential outcomes are independent of the treatment assignment conditional on the measured confounders), consistent estimation of $\tau$ follows under the {\it correct specification} of the exposure model and inclusion, in the outcome model, of all existing treatment-covariate interactions and interactions between effect modifiers and the propensity score itself. By {\it correct specification}, we mean that we measure all confounders and correctly specify the functional association between exposure and confounders \citep{shomoita2018}.

There have been different proposals to estimate causal effects following the Bayesian paradigm. \cite{mccandless2009bayesian} propose to model the propensity score and outcome using a single joint likelihood; \cite{mccandless2012adjustment} extend this joint approach to the scenario of missing confounders when external validation data are available. However, as observed in \cite{mccandless2010cutting} and \cite{zigler2013model}, the balancing property of the propensity score model may be distorted by feedback from the outcome model, which can induce bias in the causal estimators. Aiming to avoid this bias, \cite{mccandless2010cutting} propose a computational approach called ``cutting feedback'', which corresponds to a particular application of the so-called ``modularization'' technique in Bayesian inference \citep{liu2009modularization}. Its implementation assumes that the full conditional distribution of the exposure model's parameters is independent of the outcome model in each iteration of a Gibbs sampler procedure. More recently, considering  the case when there are no unmeasured confounders, \cite{Stephens2022} show that causal estimates obtained via cutting feedback methods are inherently biased due to measurement error. \cite{Stephens2022} demonstrate that the best way to follow the Bayesian specification and avoid bias in the estimation of the ATE is to follow a plug-in two-step approach. The plug-in two-step approach consists of first estimating the propensity score and then using the estimated propensity score as a known quantity in the outcome model. The estimated propensity score can be calculated, for example, using the posterior mean of the propensity score model parameters. We follow \cite{Stephens2022} and adopt a plug-in two-step method for the estimation of causal effects, while focusing on the realistic possibility of unmeasured confounding..

The remainder of this paper is organized as follows. Section \ref{sec:propMethod} describes the multilevel structure used to account for unmeasured confounding and presents the inference procedure. In Section \ref{sec:simStudy}, we provide a simulation study and discuss the effectiveness of such a multilevel approach. In Section \ref{sec:dataAnalysis}, we analyze the S\~ao Paulo TB data, and we conclude with a discussion in Section \ref{sec:discussion}.

\section{Proposed Approach}\label{sec:propMethod}

Let $S$ be a geographic region of interest, and assume that the collection of sets $S_1, \dots, S_m$ is a partition of the region $S$. Here, $S_j$ will be referred to as cluster $S_j$. Then, consider an observational study where $N = \sum_{j=1}^{m} n_j$ individuals are observed over $S$, where $n_j$ represents the number of individuals observed in cluster $S_j$. Let $Z_{ji}$ denote a binary exposure that indicates whether individual $i$ in the $j$th cluster received the treatment. Further, let $Y_{ji}$ be the outcome of interest for the $i$th individual in the $j$th cluster, and $\mathbf{X}_{ji} = ({X}_{1ji},\dots,{X}_{pji})^\top$ denotes a $p$-dimensional vector of observed covariates comprising individual and cluster-level characteristics. 

Although individual information is made available within each cluster, a key consideration for the estimation of the ATE is whether the strong ignorability assumption holds. { First, the DOT assignment is controlled by the municipal health system, which is responsible for assisting individuals that need of treatment against TB. 
	Second, \cite{paim2011brazilian} and \cite{monteiro2017use} warn about municipal politics that might impact the seeking of and access to TB treatment.} These issues raise the question of whether there is unmeasured confounding that should be accounted for when estimating the ATE under this scenario. If a propensity score approach is followed, it is natural to explore models that account for unmeasured confounding \citep{mccandless2012adjustment,reich2021review}.

The use of the propensity score in multilevel models to handle unmeasured confounding issues has been considered previously.  \cite{davis2019addressing} propose a spatial doubly robust approach that uses a multilevel structure to model the propensity score. There are two main considerations when adopting such a multilevel structure. First, the inclusion of a random effect in the propensity score model may capture unmeasured confounding structure and potentially reduce the bias of causal estimators. Second, the random structure of the multilevel model may capture an unmeasured covariate (which is not a confounder), which could impact the balance induced by the propensity score for the true (observed) confounders. The goal of this paper is to investigate the need to include a random effect in the propensity score and outcome models to account for unmeasured confounding and to then apply our findings to improve the analysis of the impact of DOT. 

\subsection{Model Specification}
Consider a binary treatment, $Z_{ji}$, such that $Z_{ji}|\delta_{ji} \sim Bernoulli(\delta_{ji})$ with
\begin{equation}
	\mbox{logit} (\delta_{ji})  = \log\left(\frac{\delta_{ji}}{1-\delta_{ji}}\right) =  \sum_{k=1}^{q} \gamma_kX_{kji} + \nu_j,
	\label{eq:exp_md}
\end{equation}
where $\delta_{ji} = P(Z_{ji}=1|\mathbf{X}_{ji})$ is the propensity score associated with the $i$th subject in cluster $j$. Further, we have that $X_{kji}$ is the value of the $k$th covariate of the $i$th subject in the $j$th cluster, with $X_{1ji} = 1$, for $i=1,\dots,n_j$ and $j=1,\dots,m$, and $\bfgamma = (\gamma_1,\dots,\gamma_q)$ a $q$-dimensional vector of regression parameters. Component $\nu_{j}$ represents a cluster-level random effect that accounts for unobserved common characteristics that individuals in cluster $j$ might share. { When considering the cluster having a random structure rather than a fixed structured (categorical observed confounder), we reduce the effective number of parameters and allow for more flexible structures based on the prior specification.}

Our analysis of the TB data presumes a binary outcome, that is $Y_{ji}=1$ if the $i$th individual in cluster $j$ obtained a diagnosis of cure, and $Y_{ji}=0$ otherwise. Thus, we assume that $Y_{ji}|Z_{ji} \sim Bernoulli(\mu_{ji})$ with
\begin{equation}
	\mbox{logit} (\mu_{ji})  = \beta_{0} + \beta_ZZ_{ji} + {B}_{ji} + \eta_j.
	\label{eq:exp_out}
\end{equation}
{The term ${B}_{ji}$ is an unknown quantity that defines how we adjust for confounding, whereas $\eta_j$ is a cluster-level random effect similar in purpose to $\nu_j$. If there is no confounding adjustment, we set ${B}_{ji}$ to zero. Alternatively, confounding adjustment directly in the outcome model is made by replacing ${B}_{ji}$ with ${B}_{ji} = \sum_{k=1}^{q} \beta_{bk}X_{kji}$. Finally, if confounding adjustment is made by adding the propensity score to the outcome model, we have ${B}_{ji} = \beta_{b}{\delta}_{ji}$.}

Because the propensity score is unknown, in the plug-in two-step approach we replace ${\delta}_{ji}$ with a point estimate of the propensity score obtained under model (\ref{eq:exp_md}). More specifically, we assume that $\widehat{\delta}_{ji} = P(Z_{ji} = 1 \mid\mathbf{X}_{ji};\widehat{\bfgamma},\widehat{\nu}_{j})$, where $\widehat{\bfgamma}$ and $\widehat{\nu}_{j}$ are posterior point estimates of ${\bfgamma}$ and ${\nu}_{j}$, respectively. { Although we chose to estimate the propensity score using a point estimate of the model parameters posterior distribution, a more general approach would be to use a point estimate of the propensity score posterior distribution.} 

The causal effect of interest is the ATE ($\tau$), defined in Equation (\ref{eq:ateTB}). An algorithm to obtain a posterior sample of the marginal posterior distribution of $\tau$ is presented in Table \ref{tab:ateALGOTIRHM}. 
The notation $\pi_N(\cdot)$ indicates the posterior distribution evaluated over $N$ sample units. Here, the integration over the $X$-distribution is performed by averaging over the observed data. 

\begin{table}[!htp]
	\centering
	\caption{Description of the algorithm to obtain a posterior sample of the marginal posterior distribution of $\tau$ achieved by empirically averaging over the observed data.}
	\begin{tabular}{l}
		\hline
		\textbf{Algorithm to obtain a posterior sample of $\tau$ with a Gibbs Sampler}\\
		\hline
		\hspace{.3cm} Define $\displaystyle\bfbeta = (\beta_0,\beta_Z,\beta_b)$, $ \displaystyle {\boldsymbol \eta} = (\eta_1,\dots,\eta_m)$ and $\displaystyle{B}_{ji} = \beta_b{B}(\mathbf{X}_{ji};\widehat{\bfgamma},\widehat{\nu}_{j})$, for all pairs $(j,i)$\\
		\hspace{.3cm} Initialize $\displaystyle (\bfbeta,{\boldsymbol \eta} )^{(0)}\sim q(\bfbeta,{\boldsymbol \eta})$\\
		\hspace{.3cm} \textbf{for} $l = 1,\dots , L$ \textbf{do}\\
		\hspace{.9cm} {Sample} $\displaystyle\bfbeta^{(l)} \sim  \pi_N(\bfbeta |{\boldsymbol \eta}^{(l-1)})$\\
		\hspace{.9cm} {Sample} $\displaystyle{\boldsymbol \eta}^{(l)} \sim  \pi_N({\boldsymbol \eta} |\bfbeta^{(l)})$\\
		\hspace{.9cm} For ${\rm z} = 0,1$, compute $\displaystyle \theta^{(l)}_{\rm z} = \frac{1}{N} \sum_{j=1}^{m} \sum_{i=1}^{n_j} E[Y_{ji}|Z_{ji}={\rm z},{B}_{ji};\bfbeta^{(l)}, \eta_j^{(l)}]$ \\
		\hspace{.9cm} {Set} $\displaystyle {\tau}^{(l)} = \theta^{(l)}_{1} - \theta^{(l)}_{0}$\\
		\hline
	\end{tabular}
	\label{tab:ateALGOTIRHM}
\end{table}

	\section{Simulation Studies}\label{sec:simStudy}
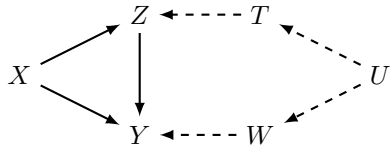
\begin{figure}[!htb]
		\centering
		\begin{tikzpicture}[scale=.8]
			[inner sep=2mm,
			place/.style={circle,draw=white!50,fill=white!20,thick},
			transition/.style={rectangle,draw=black!50,fill=black!20,thick}]
			\tikzstyle{connect}=[-latex, thick]
			
			\path ( 0,1) node (z) {${Z}$}
			(0,-1) node (y) {${Y}$}
			(4,0) node (u) {${U}$}
			(2,1) node (t) {${T}$}
			(2,-1) node (w) {${W}$}
			(-2,0) node (x) {${X}$};	
			
			\path (x) edge [connect] (z)
			(x) edge [connect] (y)
			(t) edge [connect,dashed] (z)
			(w) edge [connect,dashed] (y)
			(u) edge [connect,dashed] (t)
			(u) edge [connect,dashed] (w)
			(z) edge [connect] (y);
		\end{tikzpicture}
		\caption{DAG of the data generation mechanism for the simulation studies.}
		\label{fig:dag}
	\end{figure}

	Figure \ref{fig:dag} depicts a {\it directed acyclic graph} (DAG) to illustrate the hypothetical structure used to generate synthetic data. The dashed connections emerging from nodes ${T}$, ${W}$, and ${U}$ indicate that they are unmeasured quantities, with nodes ${T}$ and ${W}$ representing mediators in the relationship between ${Z}$ and ${U}$ and  ${Y}$ and ${U}$, respectively. Note that nodes ${T}$ and ${W}$ are independent given ${U}$, but they are not marginally independent.
	
	\noindent We assume this dependence structure and consider two scenarios. In Section \ref{sec:theoBias}, we consider both $Z$ and $Y$ to be normally distributed yielding a scenario in which bias calculations are analytically tractable. In Section \ref{sec:binaryExp}, we consider both $Z$ and $Y$ to be binary as this is commonly found in epidemiological studies and is representative of our motivating example.

	\subsection{An Analytically Tractable Example: The Linear Case}\label{sec:theoBias}
	
	In this subsection we focus on an analytical example inspired by \cite{page2017estimation} and \cite{nobre2021effects}. The idea is to construct a tractable structure that allows for analytical computation of the quantities of interest. The goal of these simulations is to investigate how different balancing score estimates affect causal estimation in the presence of measured and unmeasured confounders. Based on tractable closed forms, we analytically investigate how the inclusion of a cluster-level random effect in the propensity score model and/or the outcome model impacts causal estimates.
	
	\subsubsection*{Data generating mechanism}
	
	Let $X_{ji}$ be an individual-level covariate generated from a standard normal distribution for $j=1,\dots,m$ and $i=1,\dots,n_j$. Then, assume the following data generating mechanism for $Z$ and $Y$:
	\begin{eqnarray}
		Z_{ji} &=& \alpha_0 + \alpha_XX_{ji}  + T_{j} + \epsilon_{ji},\;\epsilon_{ji} \sim \mathcal{N}(0,\varrho^2) \label{eq:tb_Zmd-simSt-1}\\
		Y_{ji} &=& \beta_Z Z_{ji} + \beta_XX_{ji}  + W_{j} + \varepsilon_{{ji}},\;\varepsilon_{{ji}} \sim \mathcal{N}(0,\kappa^2) \label{eq:tb_Ymd-simSt-1},
	\end{eqnarray}
	where $Z_{ji}$ and $Y_{ji}$ correspond to the exposure and outcome of interest, respectively, for the $i$th unit in the $j$th cluster. Also, let $\mathbf{T} = (T_1, \dots,T_m)^\top$ and $\mathbf{W} = (W_1, \dots,W_m)^\top$ be two cluster-level covariates whose joint distribution is given by
	\begin{equation}
		\left(\begin{array}{c}
			\mathbf{T}\\ \mathbf{W}
		\end{array}\right) \sim \mathcal{N} \left[\left(\begin{array}{c}
			{\mu}_T \mathbf{1}_m\\ {\mu}_W\mathbf{1}_m
		\end{array}\right), \left(\begin{array}{cc}
			\sigma_T^2\mathbf{I}_m &\rho_{T,W}\sigma_T\sigma_W  \mathbf{I}_m\\
			\cdot & \sigma_W^2  \mathbf{I}_m
		\end{array}\right)\right],
		\label{eq:jointDist-TW}
	\end{equation}
	where $\rho_{T,W}$ is the correlation between ${\bf T}$ and ${\bf W}$ and $\sigma_T^2$ and $\sigma_W^2$ are the variances of ${\bf T}$ and ${\bf W}$, respectively. Notation ${\bf I}_m$ indicates m-dimensional identity matrix, and $\mathbf{1}_m$ indicates m-dimensional all-ones vector.
	For simplicity, quantities $\varrho$ in (\ref{eq:tb_Zmd-simSt-1}), $\kappa$ in (\ref{eq:tb_Ymd-simSt-1}),  and $\sigma_T$ and $\sigma_W$ in (\ref{eq:jointDist-TW}) are assumed to be known.
	
	\subsubsection*{Fitting the exposure model}
	
	The following linear mixed model was fitted to $Z_{ji}$'s generated according to Equation (\ref{eq:tb_Zmd-simSt-1})
	\begin{equation}
		Z_{ji} = \alpha_0 + \alpha_XX_{ji} + \nu_{j} +  \epsilon_{ji},
		\label{eq:expNormal-sim}
	\end{equation}
	such that $\nu_j$ is a latent (unobserved) component at the cluster-level, which is present to capture common structures that individuals in the same cluster $j$ might share after accounting for the covariate $X_{ji}$.
	
	Here, the balancing score that we seek to fit is the conditional expectation of $Z$ given $X$. We consider two variations of the model in Equation (\ref{eq:expNormal-sim}). First, we assume $\nu_{j} \equiv 0\;\forall_j$ and estimate $E(Z|X)$ with the best linear unbiased predictor (BLUP) of $Z$, which in this particular setting is given by $\widehat{\mathbf{BS}} = \tilde{\mathbf{X}} (\tilde{\mathbf{X}}^\top\tilde{\mathbf{X}})^{-1}\tilde{\mathbf{X}}^\top\mathbf{Z}$, where $\tilde{\mathbf{X}} = \left[{\bf 1} \mid {\bf X}\right]$ and $\mathbf{X} = (X_{11},\dots,X_{1n_1},X_{21},\dots,X_{mn_m})^\top$. The goal is to investigate estimation when the balancing score does not include an unmeasured confounder and yet it is known that such confounding exists within the data generating mechanism. Consequently, the second model attempts to capture an unmeasured confounder at the cluster-level of observations via the inclusion of a random effect; that is,  $\nu_{j}$ follows an independent zero mean normal distribution with variance $\sigma_T^2$,  $\nu_{j} \sim \mathcal{N}(0,\sigma_T^2)$. Again, we estimate $E(Z|X)$ with the BLUP of $Z$, which in this case is given by
	\[	
	\begin{split}
		\widetilde{\mathbf{BS}} &= \widehat{\alpha}_0 + \widehat{\alpha}_X\mathbf{X} + \mathbf{A}\widehat{\bfnu},
	\end{split}
	\]
	where $\mathbf{A}$ represents an $N\times m$ matrix connecting the cluster with each of its units and $\widehat{\boldsymbol{\alpha}} = (\widehat{\alpha}_0,\widehat{\alpha}_X)$ is the generalized least squares (GLS) estimator of $\bfalpha = (\alpha_0,\alpha_X)$ under the model $Z|X$ marginalized over $T$, i.e.,
	$\widehat{\boldsymbol{\alpha}} = \left[\tilde{\mathbf{X}}^\top\bfSigma_{\mathbf{Z}|\mathbf{X}}^{-1}\tilde{\mathbf{X}}\right]^{-1}\tilde{\mathbf{X}}^\top\bfSigma_{\mathbf{Z}|\mathbf{X}}^{-1}\mathbf{Z}$, with $\bfSigma_{\mathbf{Z}|\mathbf{X}} =  {\rm Var}(\mathbf{Z}|\mathbf{X})$, and $\widehat{\bfnu}$ is the BLUP of $\bfnu = (\nu_{1},\dots,\nu_{m})^\top$ in the model $[\mathbf{Z}|\mathbf{X},\widehat{\boldsymbol{\alpha}}]$. Because the covariance structure is known, $\widehat{\bfnu}$ coincides with the Bayes estimator of $\bfnu$ with a quadratic loss function (the distribution of the random effect plays the role of a prior model for $\bfnu$), and it is given by (see Chapter 2 of \cite{jiang2017asymptotic})
	\begin{equation}
		\begin{split}
			\widehat{\bfnu} &= E(\bfnu|\mathbf{Z},\mathbf{X},\widehat{\boldsymbol{\alpha}}) = \sigma_T^2\mathbf{A}^\top( \sigma_T^2\mathbf{A}\mathbf{A}^\top + \mathbf{I}_N)^{-1}(\mathbf{Z} - \widehat{\alpha}_0\mathbf{1} - \widehat{\alpha}_X\mathbf{X}).
		\end{split}
		\label{eq:random_effect_estimate}
	\end{equation}

	\subsubsection*{Fitting the outcome model}
	
	The fitted model for $Y_{ji}$ is assumed to follow
	\begin{equation}
		Y_{ji} = \beta_0 + \beta_ZZ_{ji} + B_{ji}  + \eta_{j} +  \varepsilon_{ji},
		\label{eq:outNormal-sim}
	\end{equation}
	where $B_{ji}$ indicates how the adjustment for confounding is implemented in the model; note that in the linear outcome model, if $B_{ji}$ is a propensity score correction and the propensity model is correctly specified, model \eqref{eq:outNormal-sim} leads to consistent estimation of $\beta_Z$ if the treatment-effect model is correctly specified.  Parameter $\eta_{j} \sim \mathcal{N}(0,\sigma^2_W)$ corresponds to a cluster-level random effect.
	
	Table \ref{tab:models_sim_study1} summarizes the four models under consideration for this analysis. The column $E({\bf Z}|{\bf X})$ shows the propensity score model, whereas the column $E(\mathbf{Y}|\mathbf{Z},\mathbf{X})$ describes the fitted outcome model. Models MD1 and MD2, without random effects in the outcome models, differ in the propensity score models. Models MD3 and MD4 correspond to the counterparts of MD1 and MD2, respectively, with the inclusion of the cluster-level random effect in the outcome; that is, $\eta_{j} \sim \mathcal{N}(0,\sigma^2_W)$. Models MD3 and MD4 assume that $\eta_j$ is independent of $Z_{ji}$, $X_{ji}$, and $\varepsilon_{ji}$. The outcome model, conditional on a balancing score, $\mathbf{BS}$, is then fitted as $\mathbf{Y}|\mathbf{Z},\mathbf{BS} \sim \mathcal{N}(\beta_0\mathbf{1}_N +\beta_Z\mathbf{Z} + \beta_{b}\mathbf{BS}, \boldsymbol\Sigma)$, where $\boldsymbol\Sigma = \sigma^2_W\mathbf{A}\mathbf{A}^\top + \kappa^2\mathbf{I}_N$.
	
	\begin{table}[!htb]
		\centering
		\caption{Continuous outcome simulation: fitted models. Data generated as in (\ref{eq:tb_Zmd-simSt-1})-(\ref{eq:tb_Ymd-simSt-1}). The quantities $\widehat{\mathbf{BS}}$ and $\widetilde{\mathbf{BS}}$ are the balancing scores estimated from models described in column $E({\bf Z}|{\bf X})$. The cluster-level random effects $\boldsymbol\nu$ and $\boldsymbol\eta$ are such that $\boldsymbol\nu \sim \mathcal{N}(0,\sigma^2_T{\bf I}_m)$ and $\boldsymbol\eta \sim \mathcal{N}(0,\sigma^2_W{\bf I}_m)$.}
		\begin{tabular}{l|ll}
			\hline
			\hline
			Model    & \multicolumn{1}{c}{$E({\bf Z}|{\bf X})$} & \multicolumn{1}{c}{$E(\mathbf{Y}|\mathbf{Z},\mathbf{X})$}\\
			\hline
			MD1 & $\alpha_0\mathbf{1} + \alpha_X{\bf X}$ & $\beta_0\mathbf{1}_N + \beta_Z\mathbf{Z} + \beta_{b}\widehat{\mathbf{BS}}$ \\
			MD2 & $\alpha_0\mathbf{1} + \alpha_X{\bf X} + {\bf A}\boldsymbol\nu$ & $\beta_0\mathbf{1}_N + \beta_Z\mathbf{Z} + \beta_{b}\widetilde{\mathbf{BS}}$  \\
			MD3 & $\alpha_0\mathbf{1} + \alpha_X{\bf X}$ & $\beta_0\mathbf{1}_N + \beta_Z\mathbf{Z} + \beta_{b}\widehat{\mathbf{BS}} + {\bf A}\boldsymbol\eta$ \\
			MD4 & $\alpha_0\mathbf{1} + \alpha_X{\bf X} + {\bf A}\boldsymbol\nu$ & $ \beta_0\mathbf{1}_N + \beta_Z\mathbf{Z} + \beta_{b}\widetilde{\mathbf{BS}} + {\bf A}\boldsymbol\eta$  \\
			\hline
			\hline
		\end{tabular}
		\label{tab:models_sim_study1}
	\end{table}

	In this analysis, we compare the bias of the ordinary least square (OLS) estimator of $\beta_Z$ for models MD1 and MD2, with the bias derived from the GLS estimator of $\beta_Z$ under models MD3 and MD4. Recall that if $\mathbf{Y} = \mathbf{H}\boldsymbol\beta + \boldsymbol\varepsilon$, with $\boldsymbol\varepsilon \sim \mathcal{N}(\mathbf{0},\kappa^2\mathbf{I}_N)$, the OLS estimator of $\boldsymbol\beta$ is given by $\widehat{\boldsymbol\beta} = (\mathbf{H}^\top\mathbf{H})^{-1}\mathbf{H}^\top\mathbf{Y}$. If $\mathbf{Y} = \mathbf{H}\boldsymbol\beta + \boldsymbol\varepsilon^\ast$, with $\boldsymbol\varepsilon^\ast \sim \mathcal{N}(\mathbf{0},\boldsymbol\Sigma)$ and  $\boldsymbol\Sigma$ known (and positive definite), the GLS estimator of $\boldsymbol\beta$ is given by $\widetilde{\boldsymbol\beta} = (\mathbf{H}^\top\boldsymbol\Sigma^{-1}\mathbf{H})^{-1}\mathbf{H}^\top\boldsymbol\Sigma^{-1}\mathbf{Y}$.
	
	Consider the bias in the OLS and GLS estimators of $\beta_Z$ induced by the above models when the true generating mechanism is known.
	In particular, focusing on model MD2, it can be shown (see the Appendix \ref{app:sptConf}) that
	\begin{equation}
		\begin{split}
			{\rm Bias}(\widehat{\beta}_{Z}) = \left[(\mathbf{H}^\top\mathbf{H})^{-1}\mathbf{H}^\top\left(\beta_X\mathbf{X} + \rho_{T,W}^{}\sigma_T\sigma_W\mathbf{A}\mathbf{A}^\top\bfSigma_{\mathbf{Z}|\mathbf{X}}^{-1}(\mathbf{Z} - (\alpha_0 + {\mu}_T)\mathbf{1}_N - \alpha_X\mathbf{X})\right)\right]_{(2)},
		\end{split}
		\label{eq:bias_OutMd4}
	\end{equation}
	where $\mathbf{H} = \left[\mathbf{Z}\mid \widetilde{\mathbf{BS}}\right]$ and $\left[\cdot\right]_{(2)}$ indicates the second element of the vector.
	When $\rho_{T,W} \neq 0$, ${\rm Bias}(\widehat{\beta}_{Z})$ has a complex structure; in what follows, we perform a simulation study to investigate the theoretical bias of the above models as we vary $\sigma_T^2$, $\sigma_W^2$, and $\rho_{T,W}$. We aim to show how the strength of the unmeasured confounder assumption impacts the estimation of causal effects derived from the models described in Table \ref{tab:models_sim_study1}.
	
	\subsubsection*{Results}
	
	We fix the number of clusters at $m=50$ and vary the number of units per cluster, $n$, according to the set $\{2,5,10,20\}$. The values of $\sigma_T^2$ and $\sigma_W^2$ range from $0.3$ to $3$ in increments of $0.3$, and the correlation between $T$ and $W$ ($\rho_{T,W}$) is taken from the set  $\{0,0.3,0.5\}$. For simplicity, we assume that $n_j = n$, for $j=1,\dots,m$. To complete the specification of the  data generating mechanism, we consider that $\boldsymbol{\alpha} = (1,1)^\top$, $\boldsymbol{\beta}= (\beta_Z,\beta_X)^\top = (1,1)^\top$, and $\kappa = \varrho = 1$. For each combination of $n$, $\rho_{T,W}$, $\sigma_T^2$ and $\sigma_W^2$, we simulate 1000 Monte Carlo data replicates and compute the theoretical bias of $\beta_Z$ estimates for each model presented in Table \ref{tab:models_sim_study1} (see Appendix \ref{app:sptConf} for a detailed view of the expressions). Here, we discuss the case of $(\mu_T,\mu_W)^\top=(0,0)^\top$, and results for different values can be found in Section \ref{sec:addSimSt} of the Appendix.
	
	Figures \ref{fig:case1bias} and \ref{fig:case1rmse} show heatmaps of the average absolute bias and root mean square error (RMSE), respectively, as a function of $\sigma^2_T$ and $\sigma^2_W$ obtained over 1000 Monte Carlo data replicates.
	\begin{figure}[!hbt]
		\centering
		{\includegraphics[scale=.31]{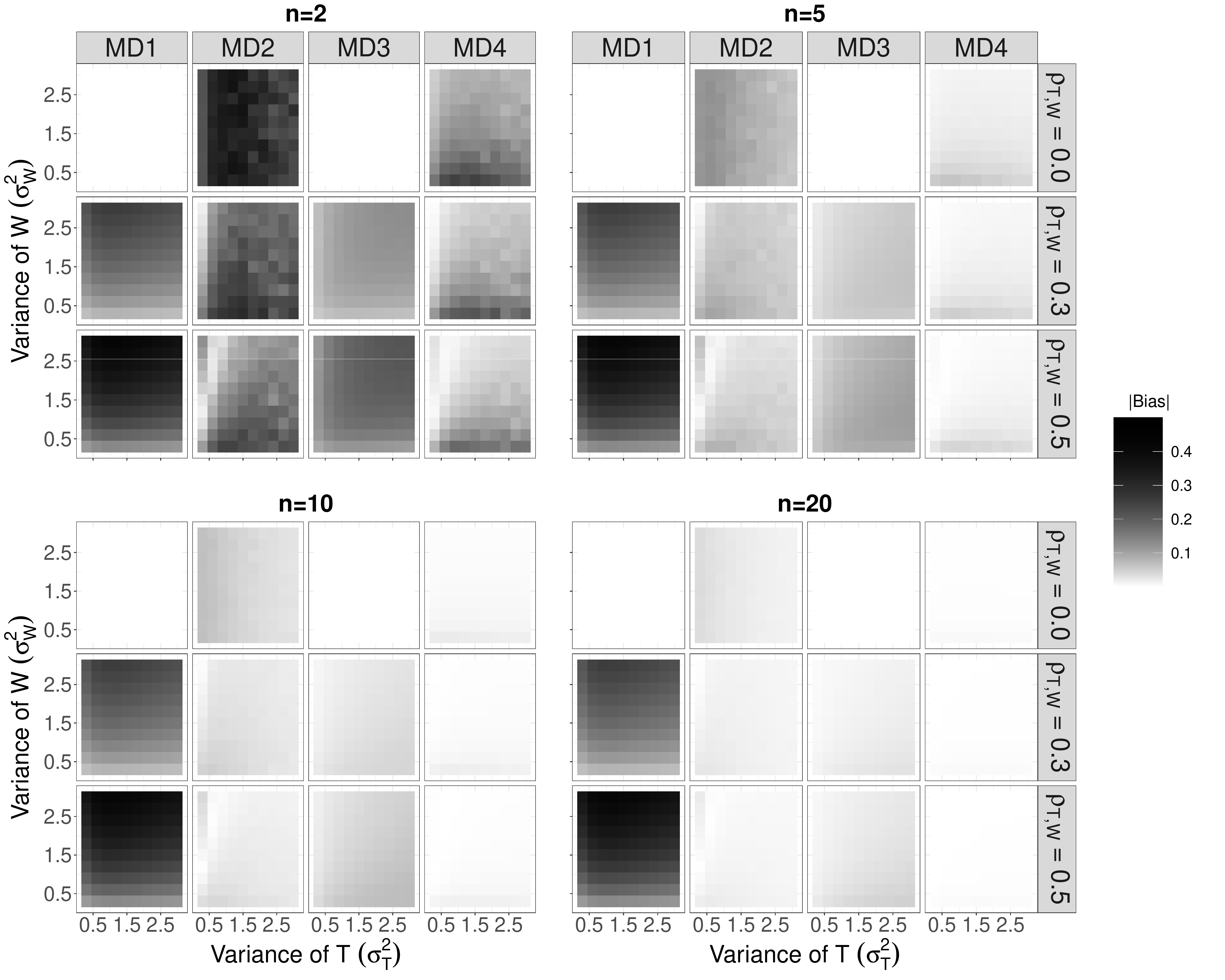}}
		\caption{Absolute bias of $\widehat{\beta}_Z$ under the models described in Table \ref{tab:models_sim_study1}. These results are averaged over 1000 Monte Carlo replicates.}
		\label{fig:case1bias}
	\end{figure}
	In particular, for each combination of $n$, $\rho_{T,W}$, $\sigma_T^2$, and $\sigma_W^2$, we compute the absolute bias for each model as $1000^{-1} \sum_{s=1}^{1000} |\widehat{\beta}_Z^{(s)} - 1|$, with $\widehat{\beta}_Z^{(s)}$ (generically) representing the GLS (or OLS) estimate of $\beta_Z$ for the $s$-th Monte Carlo replicate. Each panel pertains to a specific sample size. For each panel, the rows vary according to the value of $\rho_{T,W}$. Row 1 shows the results when $\rho_{T,W} = 0$,  Row 2 is the scenario  $\rho_{T,W} = 0.3$, and Row 3 is the scenario $\rho_{T,W} = 0.5$.
	The columns correspond to the different outcome models (moving from left to right, we have models MD1, MD2, MD3 and MD4).
	
	\begin{figure}[!hbt]
		\centering
		{\includegraphics[scale=.31]{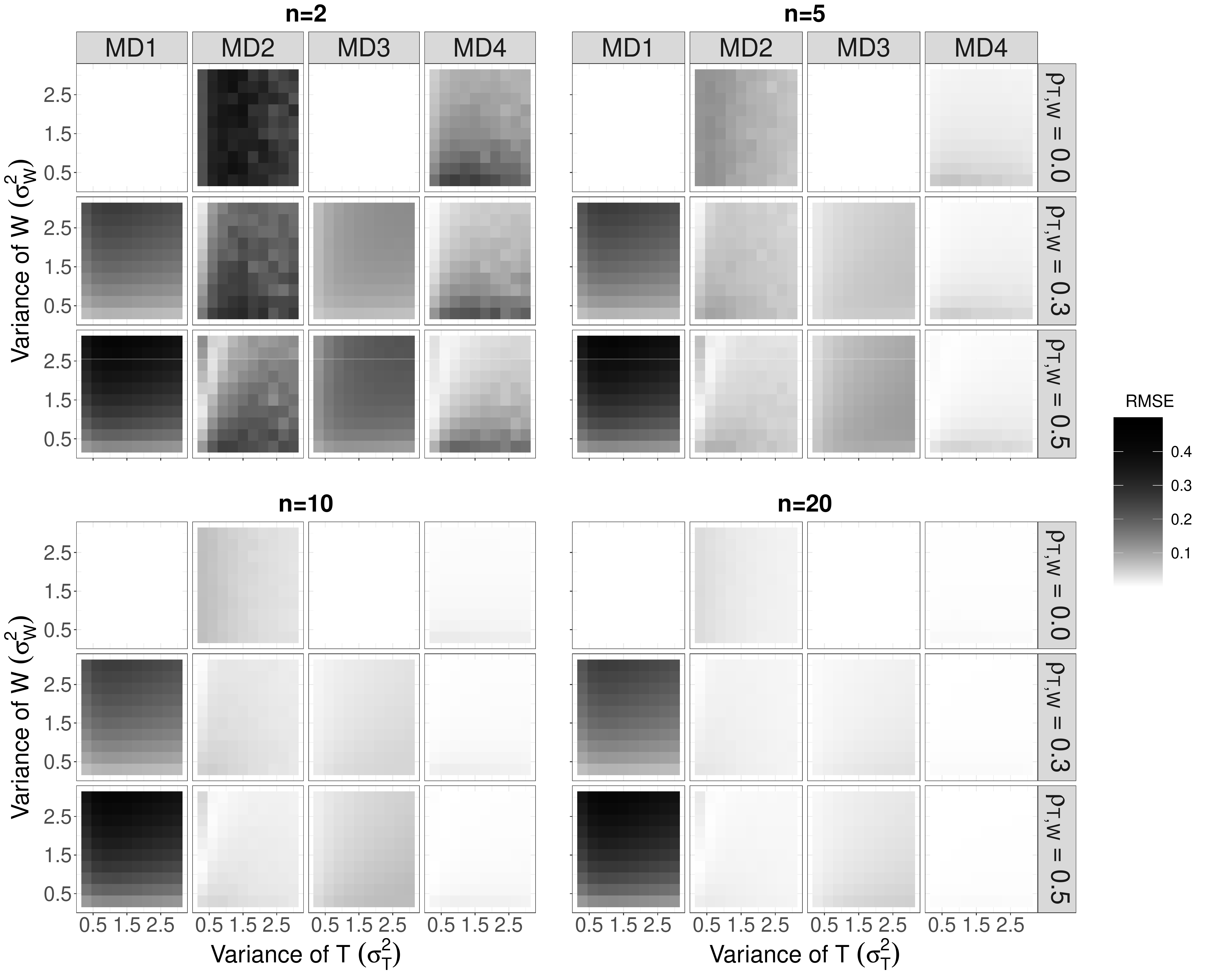}}
		\caption{RMSE of $\widehat{\beta}_Z$ under the models described in Table \ref{tab:models_sim_study1}. These results are averaged over 1000 Monte Carlo replicates.}
		\label{fig:case1rmse}
	\end{figure}

	Regardless of the sample size and relationship between $\sigma^2_T$ and $\sigma^2_W$, MD1 always provides unbiased estimation when $\rho_{T,W}=0$. However, as the correlation between $T$ and $W$ increases, so does the bias under MD1. At larger sample sizes, MD2 and MD4 provide similar results, although the former is more biased. Under MD3, for a given sample size, the bias increases with the correlation between $T$ and $W$, and as the sample size increases, the bias decreases. We also observe similarities between the results of absolute bias and RMSE, suggesting that the bias term is dominating RMSE values. In general, MD2 and MD4 show the lowest bias as the correlation between $T$ and $W$ increases.
	Further, MD2 and MD4 also show a quasi-diagonal pattern with smaller amount of bias when $\sigma_W > \sigma_T$. However, because of the complexity of the bias term, there is no clear analytical explanation for these findings. Results for different values of $(\mu_T,\mu_W)^\top$ presented in Appendix \ref{sec:addSimSt} show similar results.

\clearpage	
	
	\subsection{A Simulation Study with Binary Exposure and Outcome}\label{sec:binaryExp}
	In this section, we consider that both the exposure and outcome are binary. Therefore, we cannot rely on closed form expressions for the estimators. We follow similar ideas to those of the previous section, and the inference procedure follows the Bayesian paradigm. Samples from the posterior distributions of interest are obtained through the R package {\tt RStan} \citep{rstan}. For each data replicate, we check the convergence of the chains by considering $\widehat{R} < 1.06$ \citep{vehtari2019ranknormalization} for all parameters involved in the analysis (the $\widehat{R}$ is provided in the output of {\tt RStan}).
	
	\subsubsection*{Data generating mechanism}
	
	Initially, we consider two settings in the generation of the confounders so as to gain insights into how the distribution of  $X$ impacts the causal effect estimates. The general structure for the generation of the confounders assumes that $X_{kji} = \upsilon_{{ji}}+ \zeta_{j}$, for $k=1,2$. We consider two scenarios in the generation of $X$:
	\begin{itemize}
		\item[] Scenario 1: $\upsilon_{{ji}} \sim \mathcal{N}(0,0.1^2)$ and $\zeta_{j} \sim \mathcal{N}(0,0.4^2)$;
		\item[] Scenario 2: $\upsilon_{{ji}} \sim \mathcal{N}(0,0.25^2)$ and $\zeta_{j} \sim \mathcal{N}(0,1)$.
	\end{itemize}
	Although the total variability of $X$ is larger in Scenario 2 than in Scenario 1, the standard deviation of $\upsilon_{{ji}}$ is a quarter of the standard deviation of $\zeta_{j}$ in both settings.
	
	For each scenario, we simulate the exposure assuming that $Z_{ji}|\delta_{ji} \sim Bernoulli(\delta_{ji})$ with
	\begin{equation}
		\begin{split}
			{\rm logit}(\delta_{ji}) & = \alpha_0 + X_{1ji}\alpha_1 + X_{2ji}\alpha_2 + T_j.
		\end{split}
		\label{eq:exposure_data-simulStudy}
	\end{equation}
	For the outcome, we assume that $Y_{ji}|\mu_{{ji}} \sim Bernoulli(\mu_{{ji}})$ with
	\begin{equation}
		\begin{split}
			{\rm logit}(\mu_{{ji}}) & =  \beta_0 + X_{1ji}\beta_1 + X_{2ji}\beta_2 + X_{1ji}X_{2ji}\beta_3 + W_j.
		\end{split}
		\label{eq:binOut_data-simulStudy}
	\end{equation}
	The pair $(T_j,W_j)$ follows model (\ref{eq:jointDist-TW}) and is assumed to be independent of $X_{1ji}$ and $X_{2ji}$. Note that this specification implies a null treatment effect.
	We explore three cases in the generation of $T_j$ and $W_j$:
	\begin{itemize}
		\item[] Case 1: $T_j$ and $W_j$ independent for all $j$ $\left({\rm Corr}(T_j,W_j) = 0\right)$;
		\item[] Case 2: ${\rm Corr}(T_j,W_j) = 0.5$;
		\item[] Case 3: $T_j=W_j$.
	\end{itemize}
	Cases 1, 2, and 3 vary the strength of the unmeasured confounding, with case 1 corresponding to the setting when there is no unmeasured confounder. Cases 2 and 3 illustrate two scenarios of unmeasured confounding, with the degree of confounding in the latter being stronger than in the former.

	\subsubsection*{Fitting the exposure model}
	We explore two alternatives for the exposure model: one that only includes fixed effects in the model (PS1) and another that includes a cluster-level random effect in addition to the fixed effects (PS2). These approaches are summarized as follows:
	\begin{enumerate}
		\item[] PS1: ${\rm logit}(\delta_{ji}) = \gamma_0 + X_{1ji}\gamma_1 + X_{2ji}\gamma_2$; and
		\item[] PS2: ${\rm logit}(\delta_{ji}) = \gamma_0 + X_{1ji}\gamma_1 + X_{2ji}\gamma_2 + \nu_{j}$,
	\end{enumerate}
	where $\nu_{j} \sim \mathcal{N}(0,\varphi^2)$, for $j = 1,\dots,m$.
	
	We compute the estimated propensity score under models PS1 and PS2 using logistic regression as: $\widehat{PS}_{ji} = {\rm expit}\left(\mathbf{X}^\top_{ji}\widehat{\bfgamma}\right)$ and $\widetilde{PS}_{ji} = {\rm expit}\left(\mathbf{X}^\top_{ji}\widetilde{\bfgamma} + \widetilde{\nu}_{j}\right)$, respectively, where ${\rm expit}(a) = (1+ \exp(-a))^{-1}$. The posterior mean of $\bfgamma$ is represented by $\widehat{\bfgamma}$ under model PS1 and by $\widetilde{\bfgamma}$ under PS2; the posterior mean of $\bfnu$ under model PS2 is denoted by $\widetilde{\bfnu} = (\widetilde{\nu}_{1},\dots,\widetilde{\nu}_{m})^\top$. To complete the Bayesian specification of the exposure model and in order to be non-informative, we consider a diffuse normal prior for $\bfgamma$ and a Half-Cauchy(0,1) prior for $\varphi$ \citep{gelman2006prior} which places relatively high prior mass on small values of the parameter, but also distributes mass into a long right tail.
	
	\subsubsection*{Fitting the outcome model}
	As we aim to investigate the impact of including a cluster-level random effect in the outcome and/or exposure models when estimating the ATE, we consider a logistic regression that varies according to the inclusion of a cluster-level random effect, $\eta_{j} \sim \mathcal{N}(0,\phi^2)$, in addition to the exposure and the variants of the estimated propensity score. Table \ref{tab:models_sim_study_binOut} describes the four fitted models under consideration in this simulation study.
	
	\begin{table}[!htb]
		\centering
		\caption{Binary outcome simulation: fitted models. Data generated as in (\ref{eq:exposure_data-simulStudy})-(\ref{eq:binOut_data-simulStudy}). The quantities $\widehat{PS}_{ji}$ and $\widetilde{PS}_{ji}$ are the propensity scores estimated from models described in column ${\rm logit}(\delta_{ji})$. The cluster-level random effects $\nu_j$ and $\eta_j$ are such that $\nu_{j} \sim \mathcal{N}(0,\varphi^2)$ and $\eta_{j} \sim \mathcal{N}(0,\phi^2)$, for $j=1,\dots,m$.}
		\begin{tabular}{l|ll}
			\hline
			\hline
			Model & \multicolumn{1}{c}{${\rm logit}(\delta_{ji})$}	& \multicolumn{1}{c}{${\rm logit}(\mu_{ji})$} \\
			\hline
			MD1	 & $\gamma_0 + \gamma_1X_{1ji} + \gamma_2X_{2ji}$ & $\beta + \beta_ZZ_{ji} + \beta_{b}\widehat{PS}_{ji} $ \\
			MD2	 & $\gamma_0 + \gamma_1X_{1ji} + \gamma_2X_{2ji} + \nu_{j}$ & $\beta + \beta_ZZ_{ji} + \beta_{b}\widetilde{PS}_{ji}$ \\
			MD3	 & $\gamma_0 + \gamma_1X_{1ji} + \gamma_2X_{2ji} $ & $\beta + \beta_ZZ_{ji} + \beta_{b}\widehat{PS}_{ji} + \eta_{j}$ \\
			MD4	 & $\gamma_0 + \gamma_1X_{1ji} + \gamma_2X_{2ji} + \nu_{j}$ & $\beta + \beta_ZZ_{ji} + \beta_{b}\widetilde{PS}_{ji} + \eta_{j}$ \\
			\hline
			\hline
		\end{tabular}
		\label{tab:models_sim_study_binOut}
	\end{table}
	
	Models MD1 and MD3 include the propensity scores estimated under models PS1 and PS2 in addition to the exposure, respectively.  Models MD2 and MD4 correspond to the counterparts of models MD1 and MD3, respectively, that include an independent cluster-level random effect in addition to the exposure and propensity score. To complete the model specification, we assign a diffuse normal prior for $\bfbeta = (\beta,\beta_Z,\beta_{b})^\top$ and a Half-Cauchy(0,1) prior for $\phi$. 
	
	\subsubsection*{Results}
	
	We generate data assuming $m = 50$ and $n_j=n=4$. This results in $200$ observations.
	For all scenarios, we fixed $\bfalpha = (1,1,1)^\top$, $\bfbeta = (\beta_0,\beta_1,\beta_2,\beta_3)^\top  = (0,1/2,-1/2 ,1/4)^\top$ and $\sigma_T^2 = \sigma_W^2 = 1$. For each data generation setting, we compare the posterior mean of the ATE over 1000 Monte Carlo data replicates. Here, we assume that $(\mu_T,\mu_W)^\top = (0,0)^\top$; different settings of $(\mu_T,\mu_W)^\top$ are presented in Section \ref{sec:addSimSt2} of the Appendix.
	
	We first assess the balance induced for the measured covariates by the propensity score estimated under models PS1 and PS2. To check the balance of the covariates between treated and untreated units we use the standardized mean difference (SMD) \citep{austin2009balance}.
	In particular, we use an SMD weighted by $\omega =  {z}/ps + (1-{ z})/(1-ps)$, with $z$ corresponding to a particular value of the exposure and $ps$ referring to the propensity score; this is a metric of the balancing induced by the propensity score \citep{austin2009balance}. An absolute value less than 10\% suggests that there is no concerning imbalance between treated and untreated units \citep{austin2009using}.
	
\begin{figure}[!htb]
		\centering
		\includegraphics[scale=.45]{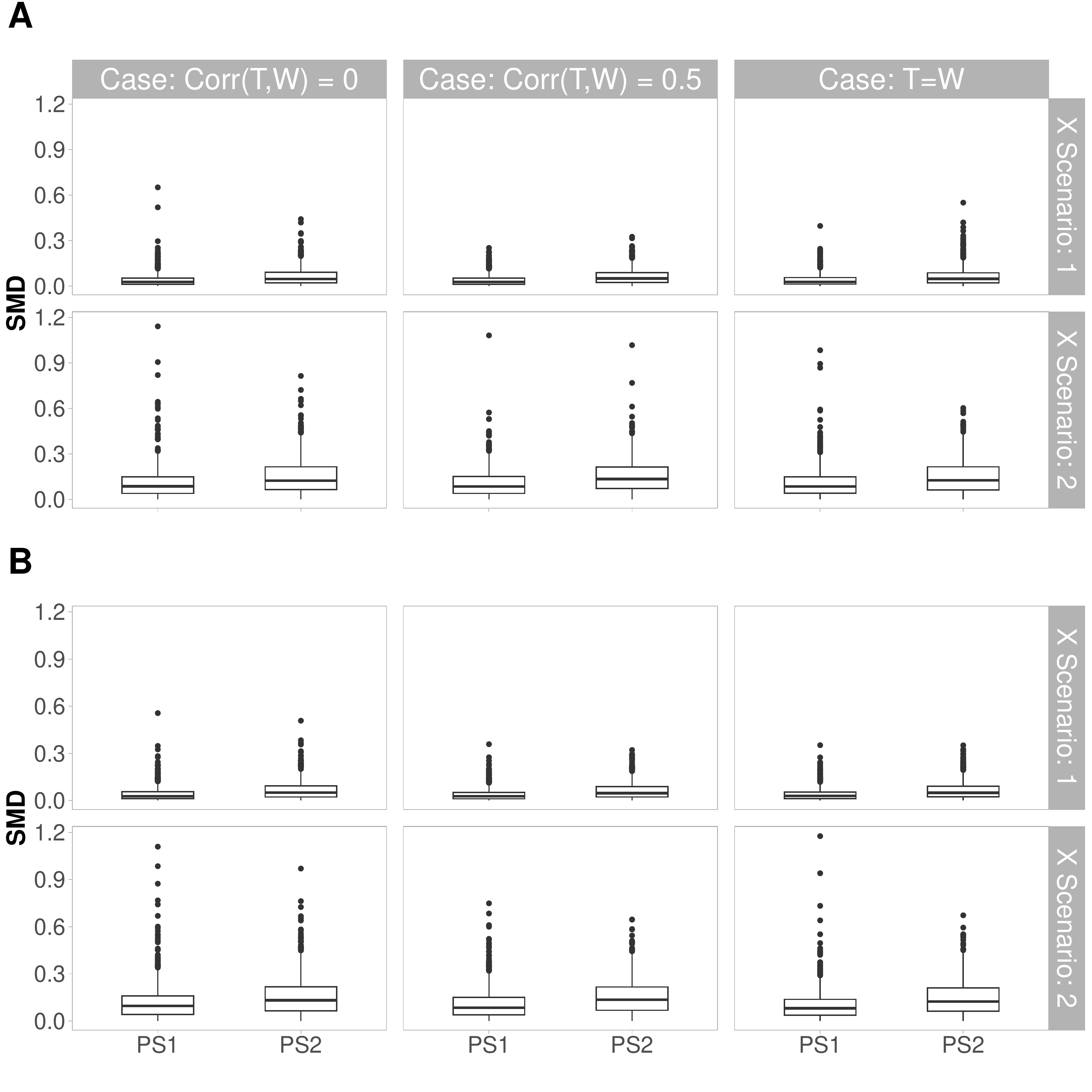}
		\caption{Boxplots of the weighted SMD of $X_1$ (Panel A) and $X_2$ (Panel B) under propensity scores estimated from models PS1 and PS2 over 1000 Monte Carlo data replicates. The columns correspond to the levels of correlation between $T$ and $W$; rows correspond to the two scenarios of X distributions in the true data generation mechanism. In the data generation model, $X_{kji} = \upsilon_{{ji}}+ \zeta_{j}$, for $k=1,2$. The labels `X Scenario: 1' and `X Scenario: 2' assume that $\upsilon_{{ji}} \sim \mathcal{N}(0,0.1^2)$ and $\zeta_{j} \sim \mathcal{N}(0,0.4^2)$, and $\upsilon_{{ji}} \sim \mathcal{N}(0,0.25^2)$ and $\zeta_{j} \sim \mathcal{N}(0,1)$, respectively.}
		\label{fig:smd}
	\end{figure}

	Figure \ref{fig:smd} shows the boxplots of the SMD associated with $X_1$ (Panel A) and $X_2$ (Panel B) evaluated for each of the 1000 Monte Carlo data replicates.
	The rows correspond to Scenario 1 and Scenario 2, respectively;
	the columns correspond to the different settings for the association between $T$ and $W$. Because the number of SMD values that exceed $10\%$ is greater for the propensity scores estimated according to model PS2, the results suggest that propensity model PS2 tends to perform more poorly than PS1. Moreover, PS2 yields greater variability in the values of the weighted SMD than PS1. Comparing Scenarios 1 (Panel A) and 2 (Panel B), there is more uncertainty in SMD values when the measured confounders variance in the population increases.

\begin{figure}[!htb]
		\centering
		\includegraphics[scale=.45]{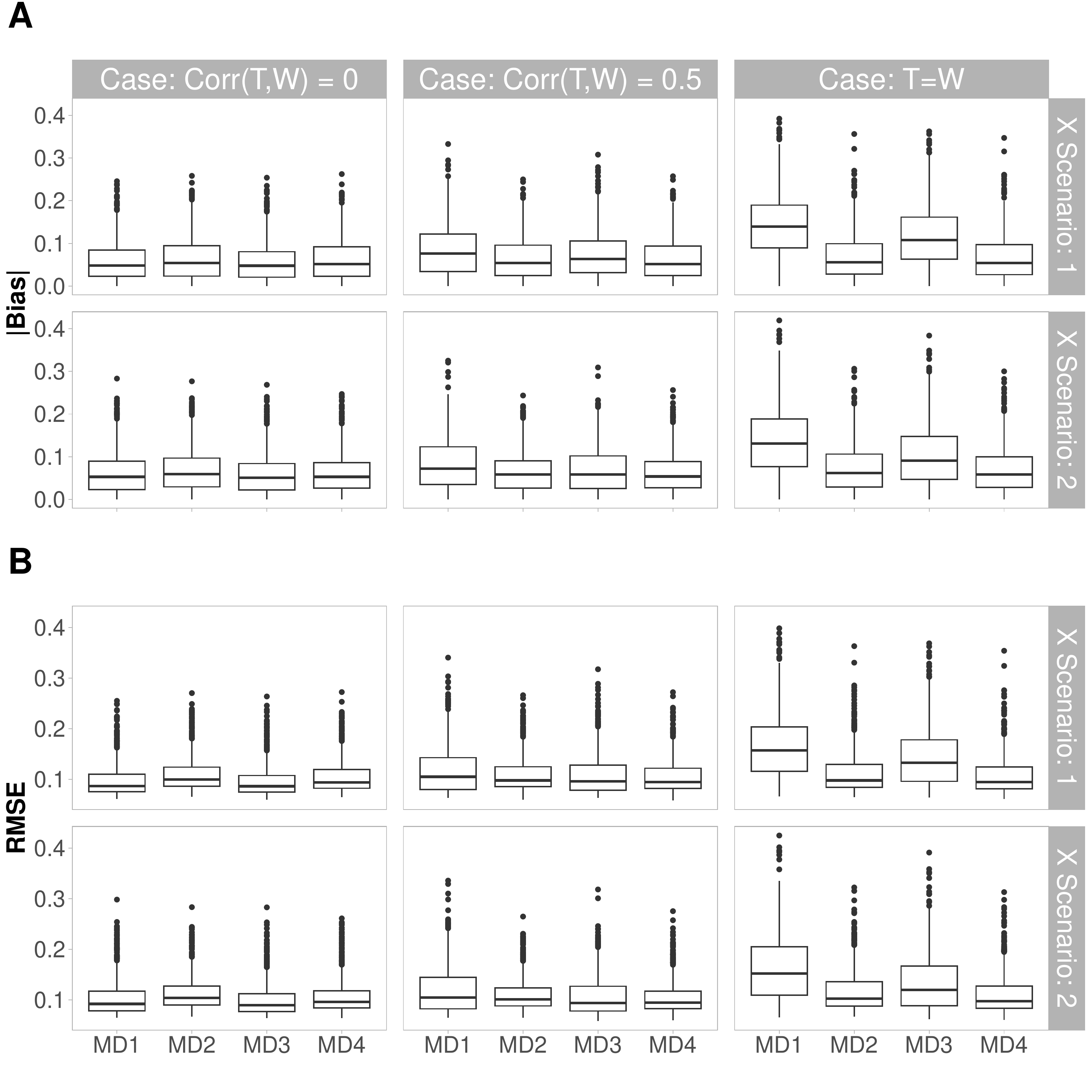}
		\caption{Boxplots of the absolute bias (Panel A) and RMSE (Panel B) of the ATE for the models described in Table \ref{tab:models_sim_study_binOut} over 1000 Monte Carlo replicates assuming binary  exposure and outcome. The columns correspond to variations in the correlation between $T$ and $W$; rows correspond to the two scenarios of X distributions in the true data generation mechanism. In the data generation model, $X_{kji} = \upsilon_{{ji}}+ \zeta_{j}$, for $k=1,2$. The labels `$X$ Scenario: 1' and `$X$ Scenario: 2' assume that $\upsilon_{{ji}} \sim \mathcal{N}(0,0.1^2)$ and $\zeta_{j} \sim \mathcal{N}(0,0.4^2)$ and $\upsilon_{{ji}} \sim \mathcal{N}(0,0.25^2)$ and $\zeta_{j} \sim \mathcal{N}(0,1)$, respectively.}
		\label{fig:simOutBinary-ate}
	\end{figure}

	Panels of Figure \ref{fig:simOutBinary-ate} display boxplots of the absolute bias and RMSE associated with the posterior mean of the ATE estimated according to the models described in Table \ref{tab:models_sim_study_binOut}. 
	More specifically, the RMSE was calculated as ${\rm RMSE}_j = \sqrt{{\rm Var}(\tau_j) + {\rm Bias}^2\left(E(\tau_j)\right)}$, where $E(\tau_j)$ and ${\rm Var}(\tau_j)$ are the posterior mean and variance of the ATE for the $j$-th Monte Carlo data replicate.
	If the strong ignorability assumption is valid (first column), models MD2 and MD4 (which condition on PS2) yield higher values of absolute bias and RMSE than models MD1 and MD3 (which condition on PS1). In contrast, if strong ignorability does not hold (second and third columns), the bias and RMSE of models MD2 and MD4 are less than that of models MD1 and MD3, particularly under the case $T=W$.
	Concerning the $X$ distribution, particularly when $T=W$, model MD3 provides results more similar to MD2 and MD4 under Scenario 2 (second row) than under Scenario 1 (first row).
	In Appendix \ref{sec:addSimSt2}, we consider variations of the mean vector of $(T_j,W_j)$ when ${\rm Corr} (T_j,W_j) = 0.5$. Results support that even in presence of confounding the conclusions are sensible to the data generating mechanism with specification of $\mu_T$ and $\mu_W$ also playing a role in this discussion.

\paragraph{Summary of the results}
When using multilevel models with a propensity score to address unmeasured confounders, it is typical to assume the validity of the latent ignorability assumption.
{This assumption states that potential outcomes are independent of the treatment assignment given a propensity score constructed based on measured confounders and random effects that accurately capture the unmeasured confounding \citep{reich2021review}.} As illustrated in Section \ref{sec:theoBias}, consistent estimation of causal effects might be achieved asymptotically when the number of units in each cluster increases.

The simulation studies suggest that the inclusion of a cluster-level random effect in the propensity score model may induce some imbalance of measured confounders between treatment groups. Consequently, when there is no unmeasured confounding, including a cluster-level random effect in the propensity score may induce poor estimation of causal effects. When investigating the inclusion of a cluster-level random effect in the outcome model, there is potential for bias reduction of causal estimates due to unmeasured confounders. We hypothesize that this bias reduction is due to the cluster-level random effect, which helps to achieve correct model specification by acting as a surrogate for the unmeasured confounder. Finally, the inclusion of a cluster-level random effect in the propensity score model can clearly reduce bias of causal effects in the presence of unmeasured confounders, as observed in Sections \ref{sec:theoBias} and \ref{sec:binaryExp}.
	
\clearpage
	
	\section{Data Analysis}\label{sec:dataAnalysis}
	
This section analyses the TB data described in Section \ref{sec:motivation}. 
Let $Y_{ji}$ be a binary variable denoting if individual $i$ in the $j$th city has a diagnosis of `cure' at the end of the treatment. Let $Z_{ji}$ be a binary variable indicating if individual $i$ in the $j$th city received DOT during the treatment of TB. Also, let $\mathbf{X}_{ji}$ be a vector of potential measured confounders including unit-level and cluster-level covariates. The individual-level characteristics include indicator variables for {Acquired Immunodeficiency Syndrome} (AIDS), diabetes, (illicit) drug user, alcoholic, homeless, gender, current prisoner, { diagnosis of mental illness} and smoker. Additionally, type of TB, 
and age (in years) are available. At the cluster-level, only the {Human Development Index} (HDI) is available, which is a proxy for the city-level socio-economic environment.

{A potential limitation to the work concerns the definition of the outcome. The original data file had ten possible outcomes recorded, which we reduced to a dichotomous measure of treatment success or failure. We opted, for example, to categorize individuals who discontinued treatment after more than $30$ days, those who died from TB, and those who experienced drug resistance as non-cured (i.e., treatment failure coded $Y_{ji}=0$). We provide full details in Section \ref{app:outDef} of the Appendix below.}

	\subsubsection*{Fitting the exposure model}
	We fit three variations of the model described in Equation (\ref{eq:exp_md}). The first model assumes that $\nu_{j} \equiv 0$, $\forall_j$; this is denoted as PS1. In the second model, we assume that the $\nu_j$'s are independent and identically distributed, that is, $\nu_{j}\sim \mathcal{N}(0,\varphi^2)$, $\forall_j$; this model is called PS2. The third model assumes that the $\nu_j$'s follow a spatial structure {\it a priori}; in particular, it is assumed that $\bfnu \sim \mathcal{N}(\mathbf{0},\varphi^2\mathbf{R}(\lambda))$, where $\mathbf{R}$ is a correlation matrix with each element following an exponential correlation function, that is, $R_{ij} = {\rm Corr}(\nu_{i},\nu_{j}) = {\rm exp}\left(-\lambda ||\bfs_i - \bfs_j|| \right)$, where $\bfs_j$ is the centroid of city $j$ (a two-dimensional vector of coordinates) and $||\cdot||$ is the Euclidean distance \citep{banerjee2014hierarchical}. We denote this model as PS3. The Bayesian specification of the propensity score model is completed after assuming a diffuse normal prior for $\bfgamma$, a Half-Cauchy(0,1) prior for $\varphi$
	\citep{gelman2006prior}; for $\lambda$, we assign a folded-normal prior whose mode provides a practical range (distance at which the correlation function drops to 5\%) at half of the maximum observed distance, and a reasonable large variance.
	
	Appendix \ref{app:smd} shows a balance diagnostic of the potential measured confounders based on the SMD. None of the weighted approaches (based on the estimated propensity score for each model) provides SMDs that exceeds 10\%. Further, the balance of the measured covariates tends to be worse under models that incorporate a random effect in the propensity score model (PS2 and PS3), similar to what was observed in the simulation study. Finally, in general, the balance diagnostic measures do not differ across the different prior specifications of the random effect.
	
	We also investigate the positivity assumption, which is satisfied when each unit has a positive chance of being assigned any value of the exposure \citep{petersen2012diagnosing}. A violation of the positivity assumption may require extrapolation that can lead to invalid conclusions. 
	To assess the overlap between the distributions of the propensity score among treated and untreated units, Figure \ref{fig:PS-overlap} in the appendix below presents boxplots of estimated propensity scores under each exposure model. There is clearly overlap for all models; however, those that include a random effect to account for unmeasured confounders result in greater separation between the propensity score distributions of the treated and untreated units.
	
\begin{figure}[!htb]
		\centering
		\includegraphics[scale=0.5]{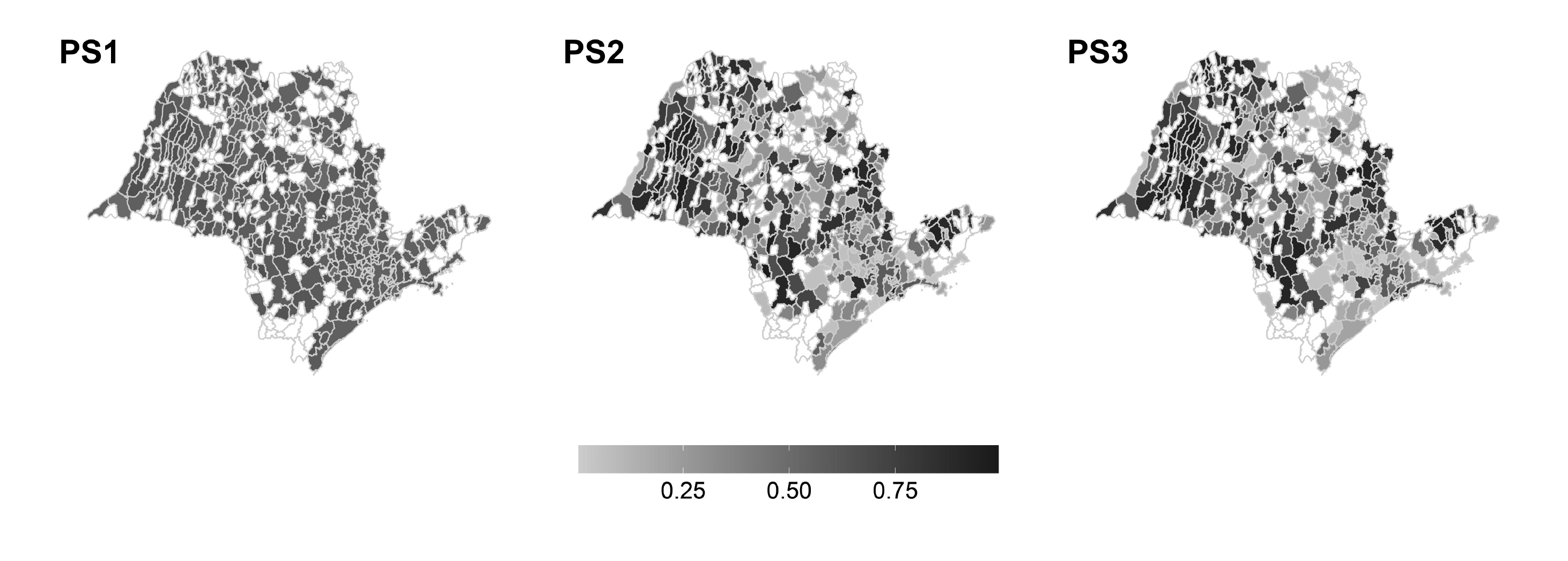}
		\caption{Map of S\~ao Paulo coloured by the estimated propensity scores averaged by city. The regions correspond to the cities of S\~ao Paulo state.}
		\label{fig:mapPS}
	\end{figure}

	Panels of Figure \ref{fig:mapPS} show the map of S\~ao Paulo coloured by the posterior means of the propensity scores averaged across the cities.
	Cities in white represent regions with no cases of TB with a concluding diagnosis in 2016. Clearly, there are clusters in the different regions of the map based on the averaged propensity scores. In particular, under PS2, groups of cities with high probabilities of treatment assignment (greater than 0.7) can be observed in the west, center, and north-east portions of the state. In contrast, there are groups of cities with low probabilities of treatment assignment (smaller than 0.3) in the north-central and south regions, suggesting that individuals with TB in these areas have lower chances of being assigned DOT than those in the west of the state, for example.
	
	\subsubsection*{Fitting the outcome model}
	Following the results in the simulation study, Table \ref{tab:modelsTB} summarizes the fifteen different models that  were fitted to the outcome.
	\begin{table}[!htb]
		\centering
		\caption{Fitted models for the diagnostic of cure at the end of TB treatment.}
		\begin{tabular}{c||l|c}
			\hline
			\hline
			Model & ${\rm logit}(P(Y_{ji} = 1|Z_{ji},\mathbf{X}_{ji}))$ & Distribution of the random effect\\
			\hline
			M1		& $\beta + Z_{ji}\beta_Z$  & $-$ \\
			M2		& $\beta + Z_{ji}\beta_Z + \eta_j $ & $\bfeta | \phi \sim \mathcal{N}(\mathbf{0},\phi^2\mathbf{I}_N)$  \\
			M3		& $\beta + Z_{ji}\beta_Z + \eta_j $ & $\bfeta | \phi,\vartheta \sim \mathcal{N}(\mathbf{0},\phi^2\mathbf{R}(\vartheta))$ \\
			M4		& $\beta + Z_{ji}\beta_Z + \mathbf{X}^\top_{ji}\bfbeta_b $ &  $-$\\
			M5		& $\beta + Z_{ji}\beta_Z + \mathbf{X}^\top_{ji}\bfbeta_b + \eta_j$ &  $\bfeta| \phi \sim \mathcal{N}(\mathbf{0},\phi^2\mathbf{I}_N)$ \\
			M6		& $\beta + Z_{ji}\beta_Z + \mathbf{X}^\top_{ji}\bfbeta_b + \eta_j$ & $\bfeta | \phi,\vartheta \sim \mathcal{N}(\mathbf{0},\phi^2\mathbf{R}(\vartheta))$\\
			M7		& $\beta + Z_{ji}\beta_Z + \mbox{PS1}_{{ji}}\beta_{b} $ & $-$\\
			M8		& $\beta + Z_{ji}\beta_Z + \mbox{PS1}_{{ji}}\beta_{b} + \eta_j$ & $\bfeta | \phi \sim \mathcal{N}(\mathbf{0},\phi^2\mathbf{I}_N)$ \\
			M9		& $\beta + Z_{ji}\beta_Z + \mbox{PS1}_{{ji}}\beta_{b} + \eta_j$ &$\bfeta | \phi,\vartheta \sim \mathcal{N}(\mathbf{0},\phi^2\mathbf{R}(\vartheta))$ \\
			M10	& $\beta + Z_{ji}\beta_Z + \mbox{PS2}_{{ji}}\beta_{b} $ & $-$\\
			M11	& $\beta + Z_{ji}\beta_Z + \mbox{PS2}_{{ji}}\beta_{b} + \eta_j$ & $\bfeta | \phi \sim \mathcal{N}(\mathbf{0},\phi^2\mathbf{I}_N)$ \\
			M12	& $\beta + Z_{ji}\beta_Z + \mbox{PS2}_{ji}\beta_{b} + \eta_j$ & $\bfeta | \phi,\vartheta \sim \mathcal{N}(\mathbf{0},\phi^2\mathbf{R}(\vartheta))$ \\
			M13	& $\beta + Z_{ji}\beta_Z + \mbox{PS3}_{{ji}}\beta_{b} $ & $-$ \\
			M14 & $\beta + Z_{ji}\beta_Z + \mbox{PS3}_{{ji}}\beta_{b} + \eta_j$ & $\bfeta | \phi \sim \mathcal{N}(\mathbf{0},\phi^2\mathbf{I}_N)$ \\
			M15 & $\beta + Z_{ji}\beta_Z + \mbox{PS3}_{{ji}}\beta_{b} + \eta_j$ & $\bfeta | \phi,\vartheta \sim \mathcal{N}(\mathbf{0},\phi^2\mathbf{R}(\vartheta))$\\
			\hline
			\hline
		\end{tabular}
		\label{tab:modelsTB}
	\end{table}
	
	Model M1 is a naive approach that considers the probability of cure at the end of the TB treatment as a function of the treatment $Z_{ji}$; its counterparts further include an independent random effect at the city-level, $\bfeta | \phi \sim \mathcal{N}(\mathbf{0},\phi^2\mathbf{I}_N)$ (M2) and a spatial structured random effect, $\bfeta | \phi,\vartheta \sim \mathcal{N}(\mathbf{0},\phi^2\mathbf{R}(\vartheta))$, at the city level (M3). Model M4 includes the baseline covariates in addition to the treatment $Z_{ji}$, and its counterparts further include city level random effects either independent, {\it a priori} (M5), or spatially structured (M6). The next models consider the conditioning on estimated propensity scores. Models M7, M8, and M9 condition on $\mbox{PS1}$ with M8 and M9 having an independent random effect and a spatial structured random effect at the city-level in addition to the propensity score, respectively. Similarly,  models M10, M11, and M12 condition instead on $\mbox{PS2}$, and models M13, M14, and M15 condition instead on $\mbox{PS3}$.
	
	\subsubsection*{Results}
	
	In Section \ref{app:tb-modelcomp} of the Appendix below, we present the values of the Watanabe-Akaike information criterion (WAIC) and the leave-one-out cross validation (LOO-CV) \citep{vehtari2017practical} for all fitted exposure and outcome models. Results from Tables \ref{tab:waic_looEXP} and \ref{tab:waic_looOUT} in the appendix below show that the inclusion of a random effect decreases substantially the values of the WAIC and LOO-CV suggesting that the inclusion of city-level random effects improves model fit. These results suggest a city-level structure in the data that is partially recovered when adding a city-level random effect in both the exposure and outcome models. However, these latent estimated characteristics are not necessarily representative of unmeasured confounders. We can only state that there is an unmeasured confounder if we can be certain of the existence of a common cause for the city-level components recovered by the city-level random effects in the exposure and outcome models.
	
\begin{figure}[!htb]
		\centering
		\includegraphics[scale=.4]{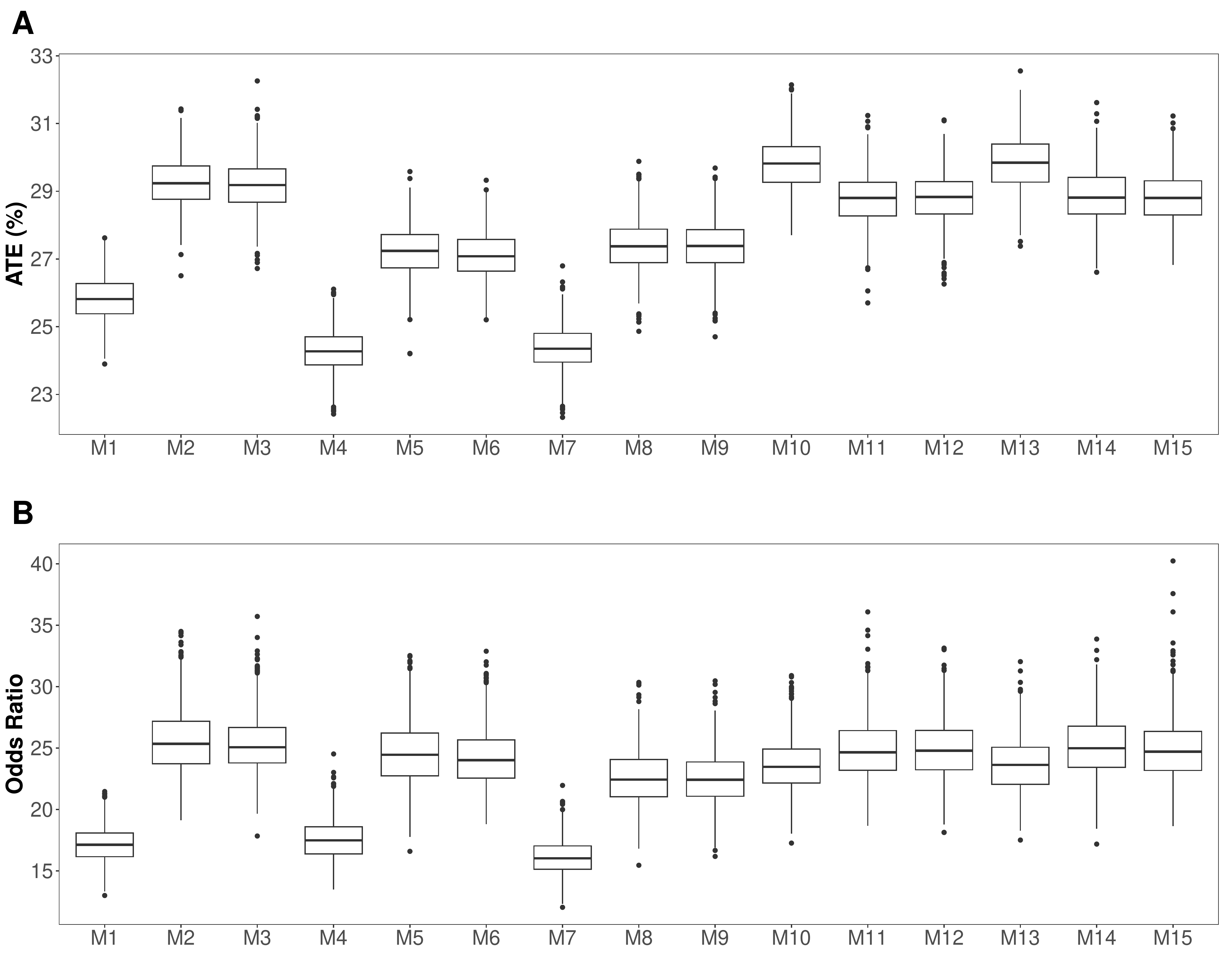}
		\caption{Posterior distributions of the ATE (Panel A) and Odds Ratio (Panel B) of the models described in Table \ref{tab:modelsTB}.}
		\label{fig:ate}
	\end{figure}

	Panels of Figure \ref{fig:ate} show boxplots of the posterior distribution of the ATE and the odds ratio of DOT for the models described in Table \ref{tab:modelsTB}. Note that, due to non-collapsibility, the ATEs and odds ratios from models that include different terms in the linear predictor (whether covariates or random effects) are not directly comparable. Although the results vary among the fitted models, all analyses indicate a large positive effect of DOT in the efficacy of TB treatment. In particular, the posterior mean of the odds ratio for model M10 is approximately $25$.
	
	Regarding the variations of such causal effects among the fitted models, we observe some interesting results. First, comparing models that only include the propensity score in addition to DOT (models M7, M10 and M13), model M7 (which does not include a random effect in the propensity score model) provides very different results from those of models M10 and M13 for both the ATE and the odds ratio.

	Focusing on models that differ only by the inclusion of a random effect in the outcome model (i.e., comparing models M1 with M2 and M3, M4 with M5 and M6, and so forth), we observe that the inclusion of a city-level random effect shifts the posterior distributions of the ATE and odds ratio to values greater than their counterparts without random effects. The exceptions are the models that condition on the propensity scores estimated under models PS2 and PS3. Further, the inclusion of the random effect changes the ATE and odds ratio estimates of models M1, M4, and M7, so that there is little overlap of the posterior distributions resulting from these models and with their counterparts that include the random effect. Clearly, the estimates obtained from models M10 and M13 are closer to their counterparts than models M1, M4, and M7 are to theirs. The literature would suggest that these changes are possibly related to issues of spatial confounding \citep{Clayton1993spatial,Paciorek2010}, minimal changes being typically associated with a lower impact of spatial confounding \citep{Reich2006,dupont2021spatial}. As shown in \cite{nobre2021effects}, in the context of multilevel structured observations, this phenomenon may also occur in settings of non-spatial structured random effects.

	\section{Discussion}\label{sec:discussion}

One of the most challenging problems in studying causal effects under observational studies occurs when the assumption of strong ignorability does not hold. Some of the current literature discusses this scenario when validation data are available \citep{mccandless2012adjustment,lin2014adjustment} or in the context of instrumental variables  \citep{angrist1996identification,adhikari2019nonparametric}. 
In this work, the proposed approach considered multilevel models in combination with propensity score regression when neither validation data nor an instrumental variable is available in the dataset. The goal was to investigate the potential for bias reduction when including a random effect in the propensity score or outcome models to account for unmeasured confounding. The discussions were motivated by the application of a multilevel model to measure the causal effect of DOT in the efficiency of the treatment against TB in Brazil.

The simulation studies of Section \ref{sec:simStudy} investigate new aspects about the effects of including random effects to account for unmeasured confounders. The results show that the inclusion of a random effect in the propensity score model cannot eliminate bias and may even introduce some bias and provide more conservative estimates when there are no unmeasured confounders. However, if the strong ignorability assumption does not hold, models that include a random effect in the model for the propensity score provide better results than their counterparts without cluster-level random effects.

In the TB data analysis, the inclusion of a random effect in the propensity score models worsened the  balance of measured confounders. However, it also improved considerably the measures of model comparison of the propensity score and outcome models (Section \ref{app:tb-modelcomp} of the Appendix). Moreover, as the outcome models are fitted conditional on {{PS2}} or {{PS3}}, the inclusion of the random effect in the outcome model does not change the ATE and odds ratio to the same degree as observed in the models that conditioned on {{PS1}}. These results suggest that the plug-in two-step approach, conditional on propensity scores models that attempt to account for unmeasured confounding via a random effect, can also be considered as an alternative to reduce confounding in the sense of spatial confounding \citep{guan2020spectral,dupont2021spatial,marques2021multivariate}. 

According to our analyses, the inclusion of a random effect in the propensity score and outcome models should be adopted when:~(1) there is reasonable belief that unmeasured confounding is present and (2) if the balance diagnostics of observed confounders are not dramatically affected by the inclusion of the cluster-level random effect in the propensity score model.

	\vspace{1cm}
	
	\paragraph*{\textit{Data Availability}:} The dataset used in the analysis is available in the GitHub webpage: \url{https://github.com/widemberg-nobre/TBdataset.git}.

	\subsection*{Acknowledgments}
	
	WSN was supported by awards from the Conselho Nacional de Desenvolvimento Científico e Tecnológico (CNPq), Brazil (Scholarship 140529/2017-9), and Funda\c{c}{\~a}o de Amparo à Pesquisa do Estado do Rio de Janeiro (FAPERJ), Brazil (Scholarship E-26/200.809/2019). AMS, EEMM, and DAS are all supported by individual Discovery Grants from the Natural Sciences and Engineering Research Council of Canada (NSERC). EEMM is also supported by a career award from the Fonds de recherche du Qu\'ebec - Sant\'e and a Canada Research Chair. WSN was also funded by the Emerging Leaders in the Americas Program, with the support of the Government of Canada. The authors would like to thank Daniele M. Pelissari, Denise Arakaki, Barbara Reis-Santos and Patricia B. Oliveira from the Brazilian Health Ministry for all the support they provided on accessing and understanding the TB data.

	\bigskip
	

	\bibliographystyle{rss}
	
	\bibliography{BibReferences}
	
	\clearpage
	
	\appendix
	\setcounter{table}{0}
	\setcounter{figure}{0}
	\setcounter{equation}{0}
	\renewcommand\thefigure{\thesection.\arabic{figure}}
	\renewcommand\thetable{\thesection.\arabic{table}}
	\renewcommand\theequation{\thesection.\arabic{equation}}
	
	\section{Complementary Analyses for Section \ref{sec:simStudy}}	
	This section present supplementary results for the analyses of Sections \ref{sec:theoBias} and \ref {sec:binaryExp}. 
		
	\subsection{Calculations for the Simulation Study presented in Section \ref{sec:theoBias} }\label{app:sptConf}
	Let the estimator of $\beta_Z$ be denoted as
	\begin{equation}
		\widehat{\beta} = \mathbf{G}\mathbf{Y},
		\label{eq:app_beta.hat}
	\end{equation}
	where $\mathbf{G}$ is a known $3\times N $ matrix that depends on $\mathbf{Z}$ and $\mathbf{X}$.
	Because $\mathbf{G}$ is known, it follows then:
	\[
	\begin{split}
		E(\widehat{\beta}|\mathbf{Z},\mathbf{X}) &= \mathbf{G}E(\mathbf{Y}|\mathbf{Z},\mathbf{X}).
	\end{split}
	\]
	
	Thus, we have the conditional expectation $E(\mathbf{Y}|\mathbf{Z},\mathbf{X})$ given by
	\[
	\begin{split}
		E(\mathbf{Y}|\mathbf{Z},\mathbf{X}) &= \beta_Z\mathbf{Z} + \beta_X\mathbf{X} + \mathbf{A}E(\mathbf{W}|\mathbf{Z},\mathbf{X}).
	\end{split}
	\]
	
	From models (\ref{eq:tb_Zmd-simSt-1}) and (\ref{eq:jointDist-TW}) and assuming that $(\mathbf{T},\mathbf{W})$ and $\mathbf{X}$ are marginally independent, the joint conditional distribution of $(\mathbf{T},\mathbf{W},\mathbf{Z}) \mid \mathbf{X}$ is such that
	\[
	\left(\begin{array}{c}
		\mathbf{T}\\ \mathbf{W}\\ \mathbf{Z}
	\end{array}\right)  \mid \mathbf{X} \sim \mathcal{N} \left\{\left(\begin{array}{c}
		{\mu}_T\mathbf{1}_m\\ {\mu}_W\mathbf{1}_m\\ (\alpha_0 + {\mu}_T)\mathbf{1}_N + \alpha_X\mathbf{X}
	\end{array}\right), \left(\begin{array}{ccc}
		\sigma_T^2\mathbf{I}_m & \rho_{T,W}^{}\sigma_T\sigma_W  \mathbf{I}_m & \sigma_T^2\mathbf{I}_m\mathbf{A}^\top\\
		\cdot & \sigma_W^2  \mathbf{I}_m & \rho_{T,W}^{}\sigma_T\sigma_W\mathbf{I}_m\mathbf{A}^\top\\
		\cdot & \cdot & \sigma_T^2\mathbf{A}\mathbf{A}^\top+ \mathbf{I}_N
	\end{array}\right)\right\}.
	\]
	
	As a result, it follows that
	\[
	E(\mathbf{W} \mid \mathbf{Z},\mathbf{X}) = {\mu}_W\mathbf{1}_m +  \rho_{T,W}^{}\sigma_T\sigma_W\mathbf{A}^\top\left(\sigma_T^2\mathbf{A}\mathbf{A}^\top + \mathbf{I}_N\right)^{-1}(\mathbf{Z} - (\alpha_0 + {\mu}_T)\mathbf{1}_N - \alpha_X\mathbf{X}),
	\]
	and
	\[
	{\rm Var}(\mathbf{W} \mid \mathbf{Z},\mathbf{X}) = \sigma_W^2 \left\{ \mathbf{I}_m - \rho_{T,W}^{2}\sigma_T^2\mathbf{A}^\top\left(\sigma_T^2\mathbf{A}\mathbf{A}^\top + \mathbf{I}_N\right)^{-1}\mathbf{A}\right\}  .
	\]
	Finally, we have
	\[
	\begin{split}
		E(\mathbf{Y}|\mathbf{Z},\mathbf{X}) &= \beta_Z\mathbf{Z} + \beta_X\mathbf{X} + {\mu}_W\mathbf{1}_N +  \rho_{T,W}^{}\sigma_T\sigma_W\mathbf{A}\mathbf{A}^\top\bfSigma_{\mathbf{Z}|\mathbf{X}}^{-1}(\mathbf{Z} - (\alpha_0 + {\mu}_T)\mathbf{1}_N - \alpha_X\mathbf{X} )\\
		{\rm Var}(\mathbf{Y}|\mathbf{Z},\mathbf{X}) &= \kappa^2 \mathbf{I}_N + \sigma_W^2 \mathbf{A}\left\{ \mathbf{I}_m - \rho_{T,W}^{2}\sigma_T^2\mathbf{A}^\top\left(\sigma_T^2\mathbf{A}\mathbf{A}^\top + \mathbf{I}_N\right)^{-1}\mathbf{A}\right\}\mathbf{A}^\top.
	\end{split}
	\]

	If $\mathbf{G}$ is the projection matrix of the OLS (or GLS) estimator that pre-multiplies $\mathbf{Y}$, for each of the models presented in Table \ref{tab:models_sim_study1}, then we have $\mathbf{G}\mathbf{1}_N = (\;1\; , \; 0\;, \; 0\;)^\top$ and $\mathbf{G}\mathbf{Z} = (\;0\; , \; 1\;, \; 0\;)^\top$. Therefore, the bias and variance of $\widehat{\beta}$, defined in (\ref{eq:app_beta.hat}), has the general form:
	\begin{equation}
		\begin{split}
			{\rm Bias}(\widehat{\beta}_{Z}) &=\left[\mathbf{G}\left(\beta_X\mathbf{X} + \rho_{T,W}^{}\sigma_T\sigma_W\mathbf{A}\mathbf{A}^T\bfSigma_{\mathbf{Z}|\mathbf{X}}^{-1}(\mathbf{Z} - (\alpha_0 + {\mu}_T)\mathbf{1}_N - \alpha_X\mathbf{X})\right)\right]_{(2)} \\
			{\rm Var}(\widehat{\beta}_{Z}) &= \left[{\rm diag}\left\{\mathbf{G}\left(\kappa^2 \mathbf{I}_N + \sigma_W^2 \mathbf{A}\left( \mathbf{I}_m - \rho_{T,W}^{2}\sigma_T^2\mathbf{A}^\top\left(\sigma_T^2\mathbf{A}\mathbf{A}^\top + \mathbf{I}_N\right)^{-1}\mathbf{A}\right)\mathbf{A}^\top\right) \mathbf{G}^\top\right\}\right]_{(2)},
		\end{split}
		\label{eq:app_bias_betaZ}
	\end{equation}
	with $\left[\cdot\right]_{(2)}$ denoting the second element of the vector. From Equation (\ref{eq:app_bias_betaZ}), we can easily derive the bias of $\widehat{\beta}_{Z}$ for each of the models described in Table \ref{tab:models_sim_study1}.
	
	\clearpage

\subsection{Additional Simulation Analyses for Section \ref{sec:theoBias}}\label{sec:addSimSt}

\paragraph{Variations on $(\mu_T,\mu_W)^\top$}
First, we examine additional results for the tractable structure in Section \ref{sec:theoBias}. This analysis aims to evaluate the effect of estimates based on variations of $(\mu_T,\mu_W)^\top$, defined in (\ref{eq:jointDist-TW}), while holding other quantities constant as described in Section \ref{sec:theoBias}.

\begin{figure}[!htb]
	\centering
	\includegraphics[scale=0.23]{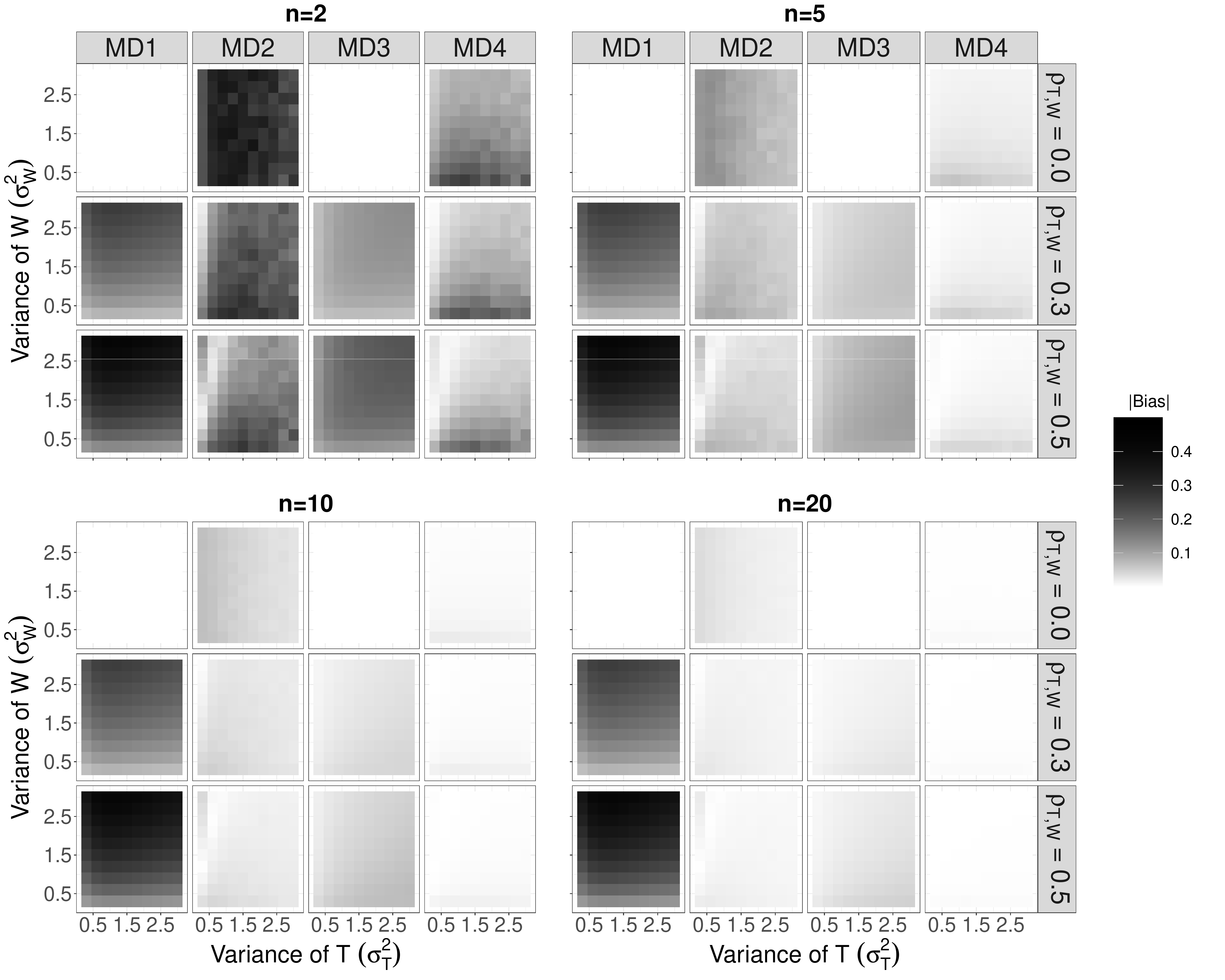}
	\caption{Absolute bias of $\widehat{\beta}_Z$ under the models described in Table \ref{tab:models_sim_study1} and assuming that $(\mu_T,\mu_W)^\top = (1,-1/2)^\top$.}
	\label{fig:bias1app}
\end{figure}

\begin{figure}[!htb]
	\centering
	\includegraphics[scale=0.23]{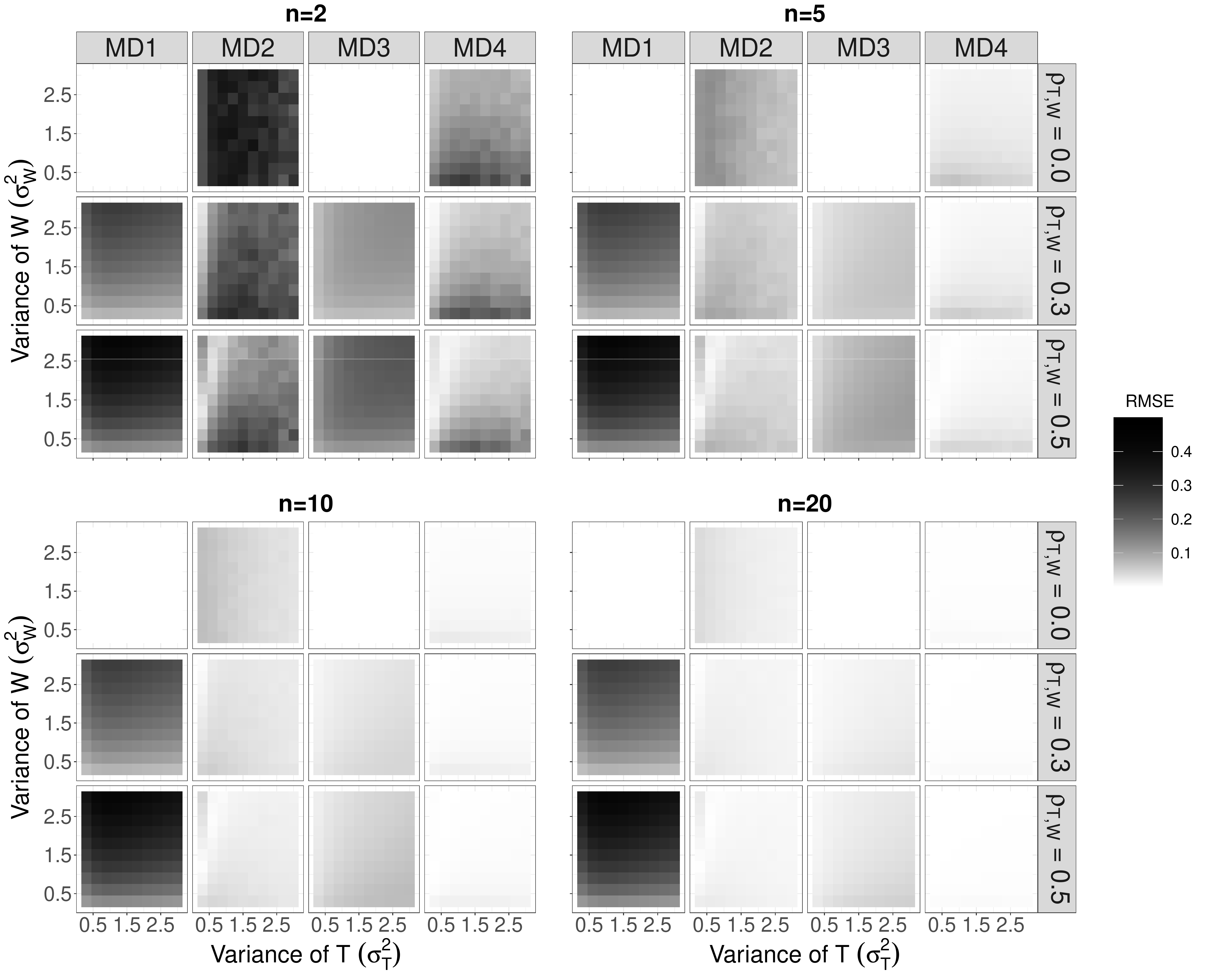}
	\caption{RMSE of $\widehat{\beta}_Z$ under the models described in Table \ref{tab:models_sim_study1} and assuming that $(\mu_T,\mu_W)^\top = (1,-1/2)^\top$.}
	\label{fig:rmse1app}
\end{figure}

\begin{figure}[!htb]
	\centering
	\includegraphics[scale=0.23]{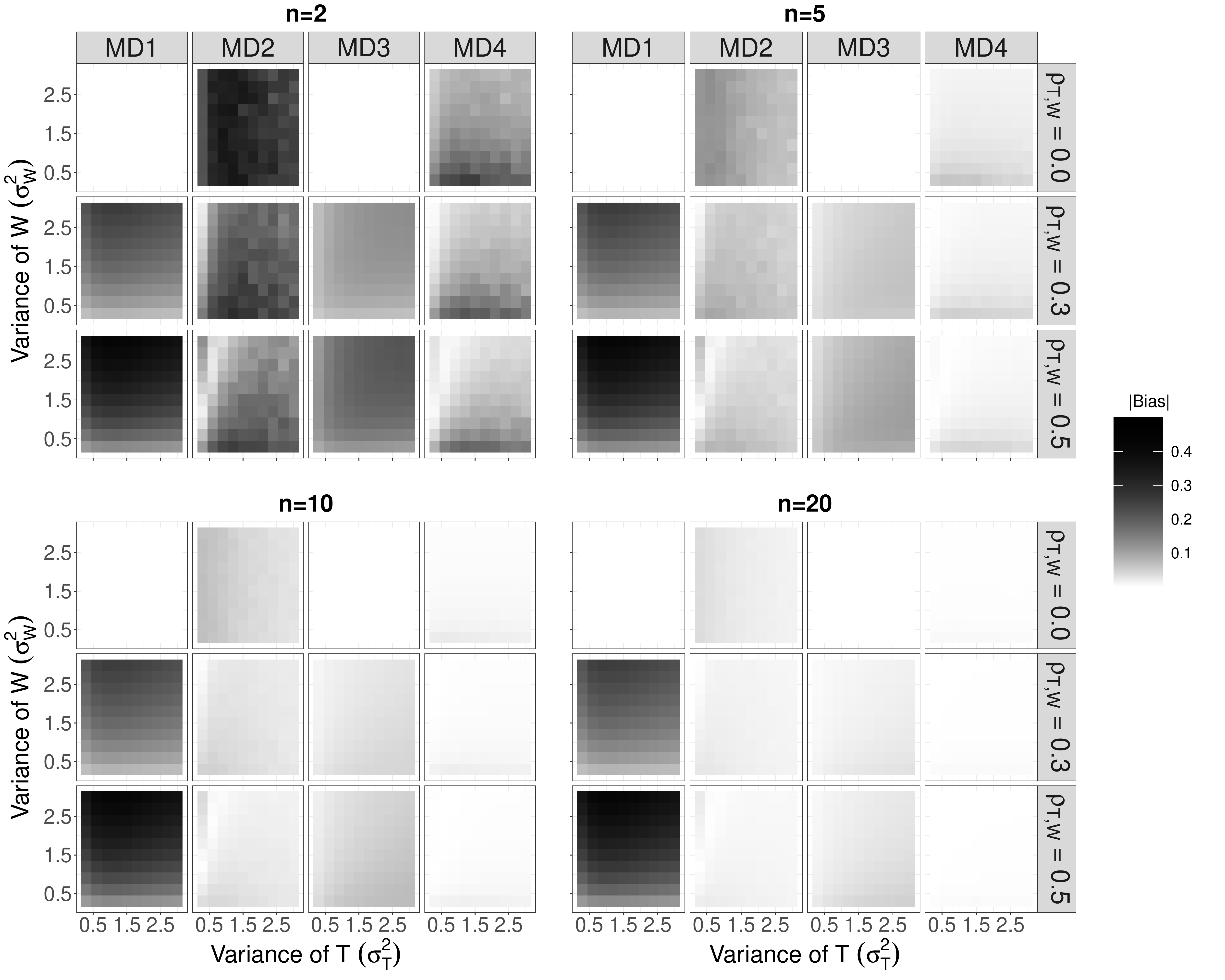}
	\caption{Absolute bias of $\widehat{\beta}_Z$ under the models described in Table \ref{tab:models_sim_study1} and assuming that $(\mu_T,\mu_W)^\top = (-1/2,1)^\top$.}
	\label{fig:bias2app}
\end{figure}

\begin{figure}[!htb]
	\centering
	\includegraphics[scale=0.23]{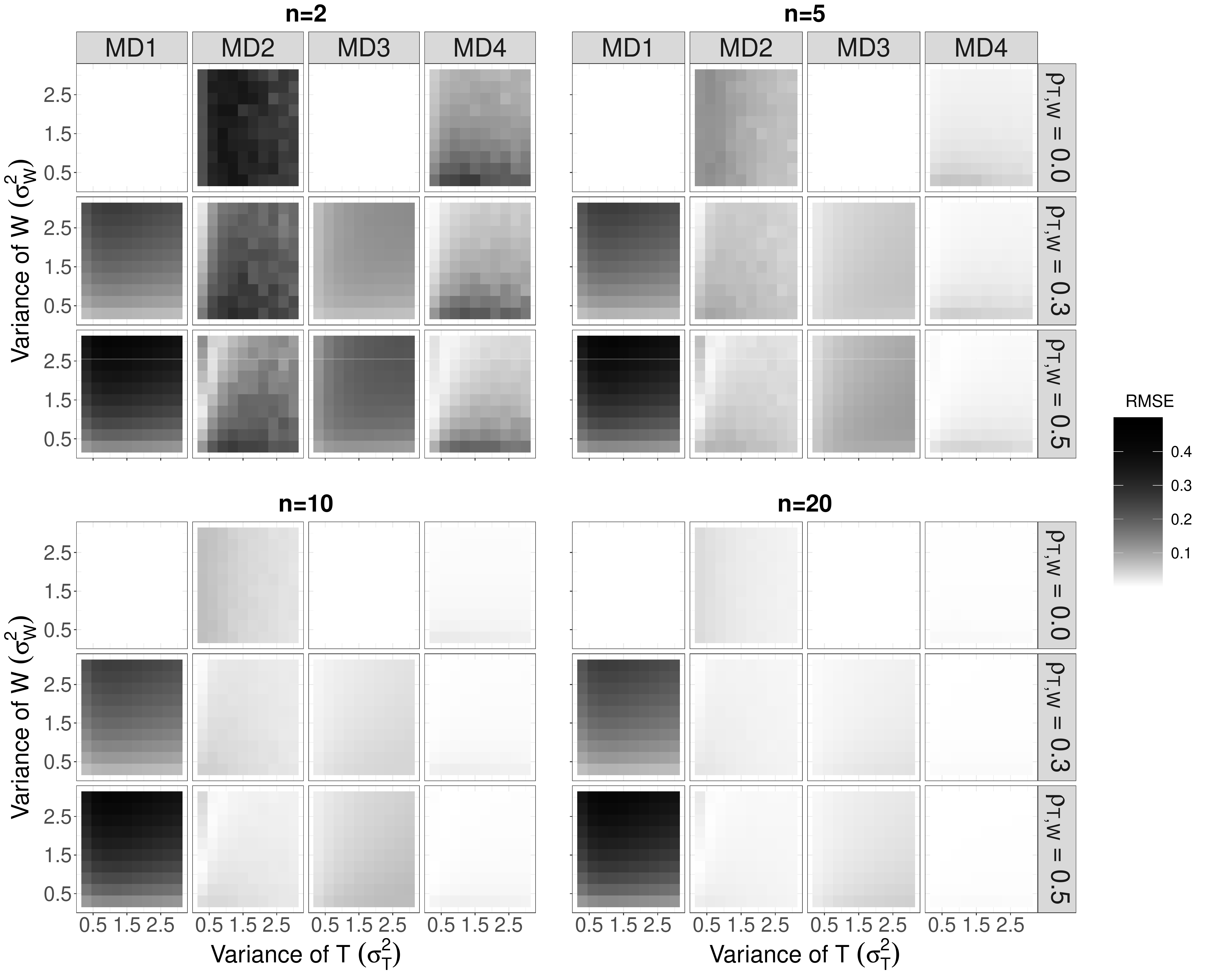}
	\caption{RMSE of $\widehat{\beta}_Z$ under the models described in Table \ref{tab:models_sim_study1} and assuming that $(\mu_T,\mu_W)^\top = (-1/2,1)^\top$.}
	\label{fig:rmse2app}
\end{figure}

Figures \ref{fig:bias1app}-\ref{fig:rmse2app} present heatmaps
of the average absolute bias and root mean square error (RMSE), respectively, for $(\mu_T,\mu_W)^\top = (1,-1/2)^\top$ (Figures \ref{fig:bias1app} and \ref{fig:rmse1app}) and $(\mu_T,\mu_W)^\top = (-1/2,1)^\top$ (Figures \ref{fig:bias2app} and \ref{fig:rmse2app}). The results are displayed as a function of $\sigma^2_T$, $\sigma^2_W$, sample size ($n$), and the correlation between $T$ and $W$, similar to the analysis in Section \ref{sec:theoBias}.


Overall, the results for both variations of $(\mu_T,\mu_W)^\top$ are similar. As previously discussed in the main text, absolute bias and RMSE results indicate that bias dominates RMSE. When $\rho_{T,W}=0$, MD1 and MD3 estimates are unbiased, while MD2 and MD4 estimates exhibit biases that decrease as $n$ increases. Conversely, as the correlation between $T$ and $W$ increases, so does the bias for MD1 and MD3. Meanwhile, MD2 and MD4 generally have lower bias and RMSE compared to their counterparts without the random effect in the propensity score model.

\clearpage

\paragraph{Outcome intercept known (equal to zero) and $(\mu_T,\mu_W)^\top = (0,0)^\top$}
Next, we examine the impact of specifying (correctly) a zero-intercept in the outcome model. Our aim is to determine if the outcome intercept estimation affects the pattern of bias and RMSE. All other values, including $(\mu_T,\mu_W)^\top$, are fixed as described in Section \ref{sec:theoBias}

\begin{figure}[!htb]
	\centering
	\includegraphics[scale=0.23]{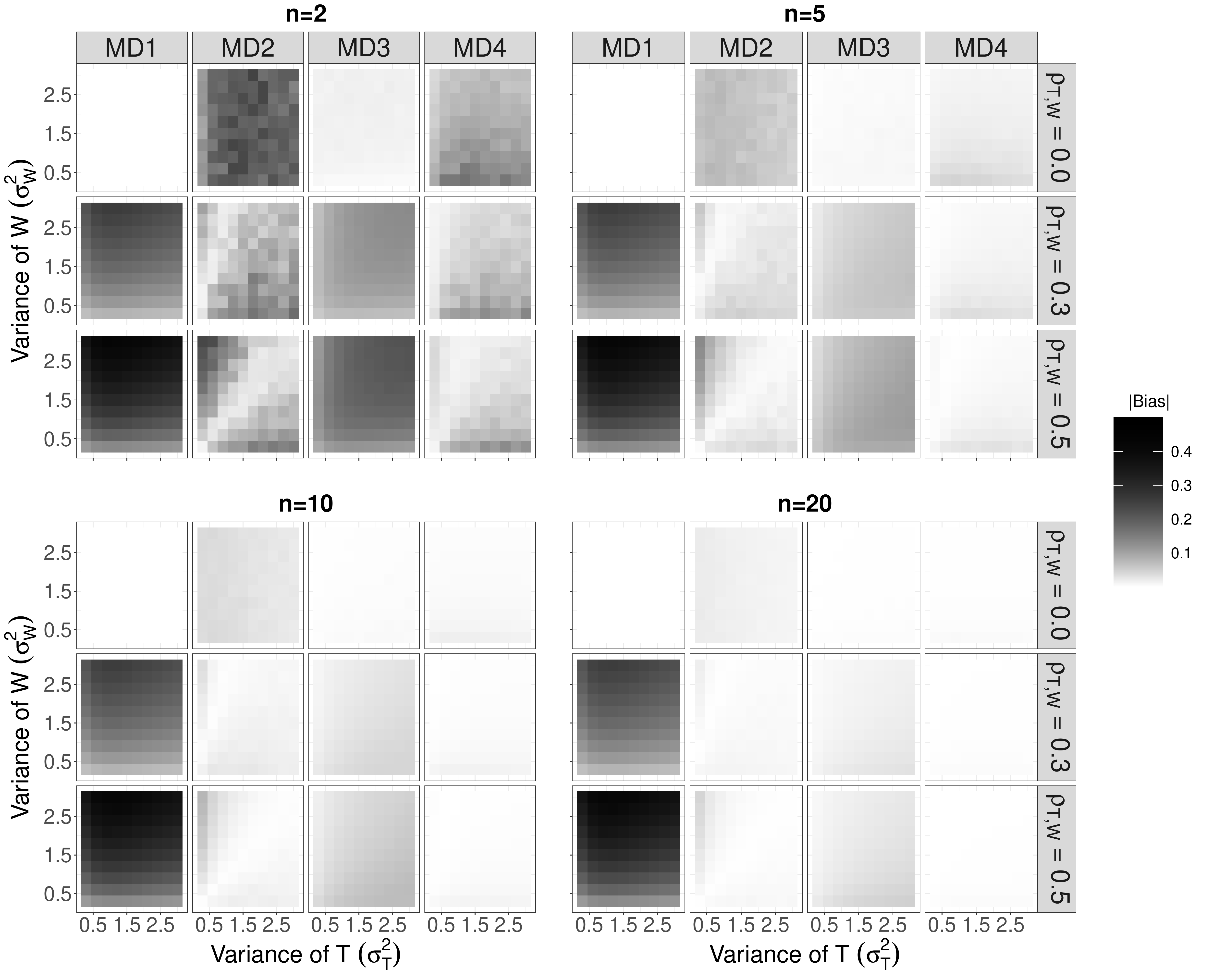}
	\caption{Absolute bias of $\widehat{\beta}_Z$ under the models described in Table \ref{tab:models_sim_study1} and assuming that $(\mu_T,\mu_W)^\top = (0,0)^\top$. The analysis involves a zero-intercept specification of the outcome model. }
	\label{fig:case1bias2app}
\end{figure}

\begin{figure}[!htb]
	\centering
	\includegraphics[scale=0.23]{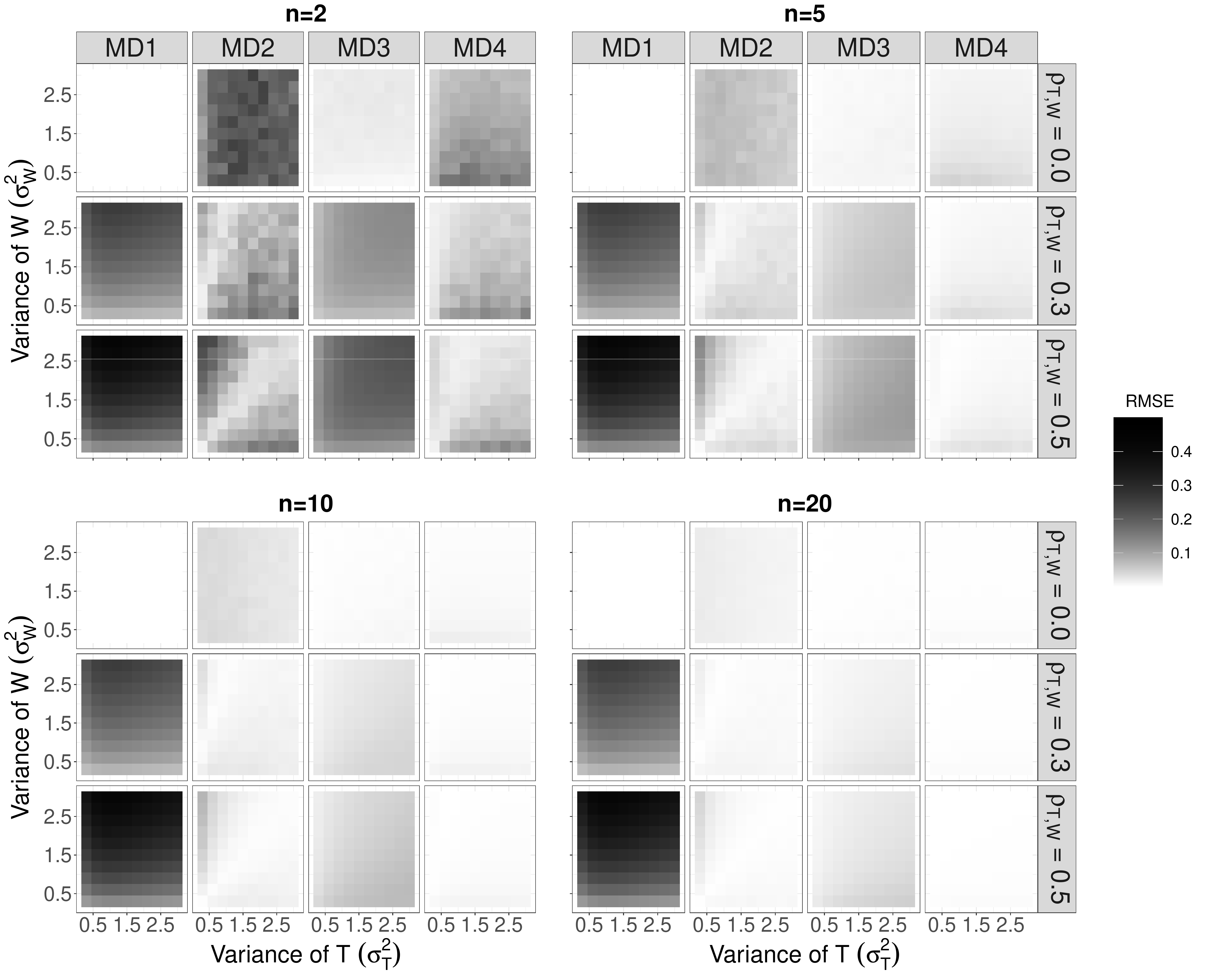}
	\caption{RMSE of $\widehat{\beta}_Z$ under the models described in Table \ref{tab:models_sim_study1} and assuming that $(\mu_T,\mu_W)^\top = (0,0)^\top$. The analysis involves a zero-intercept specification of the outcome model.}
	\label{fig:case1rmse2app}
\end{figure}

Figures \ref{fig:case1bias2app} and \ref{fig:case1rmse2app} depict heatmaps of the average absolute bias and root mean square error (RMSE), respectively, that vary as a function of $\sigma^2_T$ and $\sigma^2_W$, and considering variations of the sample size ($n$) and the correlation between $T$ and $W$.
In general, MD1 and MD3 provide results similar to the previous analyses. Further, comparisons among models, considering variation of the sample size and the correlation between $T$ and $W$, also demonstrate similar conclusions.

In general, the results from MD1 and MD3 are consistent with the previous analyses. Further, comparisons related to variations of the sample size and correlation between $T$ and $W$ reach similar conclusions.
On the other hand, MD2 and MD4 exhibit slightly different patterns than that observed in previous analyses.
Because of the complexity of the bias term, we do not have a clear analytical explanation for this result; however, the magnitude of the difference is fairly small.

\clearpage

\subsection{Additional Simulation Analyses for Section \ref{sec:binaryExp}}\label{sec:addSimSt2}
In this section, we extend the analysis of Section \ref{sec:binaryExp} by investigating different values of $(\mu_T,\mu_W)^\top$ in the data generating mechanism. The focus is on Case 2, where ${\rm Corr}(T_j,W_j) = 0.5$, with all other aspects of the simulation remaining the same as described in Section \ref{sec:binaryExp}.

\begin{figure}[!htb]
	\centering
	\includegraphics[scale=.35]{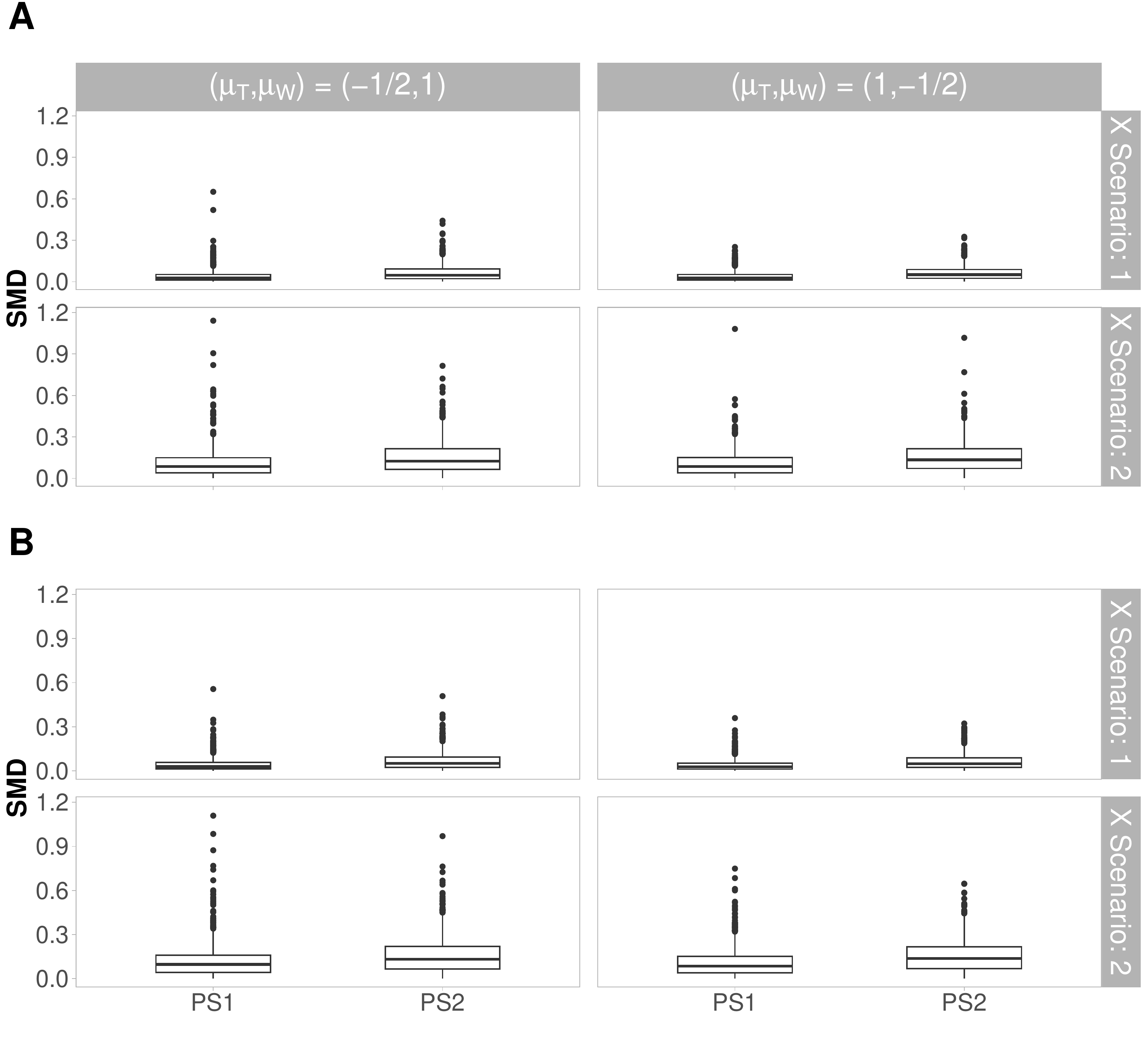}
	\caption{Boxplots of the weighted SMD of $X_1$ (Panel A) and $X_2$ (Panel B) under propensity scores estimated from models PS1 and PS2 over 1000 Monte Carlo data replicates. The columns correspond to variations of $\mu_T$ and $\mu_W$; rows correspond to the two scenarios of $X$ distributions in the true data generation mechanism. We considered ${\rm Corr}(T_j,W_j)=0.5$.}
	\label{fig:smd2app}
\end{figure}

Panels of Figure \ref{fig:smd2app} display boxplots of the SMD values of $X_1$ (Panel A) and $X_2$ (Panel B)
for 1000 Monte Carlo replicates. The columns correspond to variations of $(\mu_T,\mu_W)$; rows correspond to variations of the $X$-distribution in the true data generation mechanism. Results are consistent with those in Section \ref{sec:binaryExp}, showing that PS2 performs worse than PS1 with more SMD values exceeding 10\%. Additionally, the results indicate that the variance of the $X$'s has a significant impact on the uncertainty of the SMD value.

\begin{figure}[!htb]
	\centering
	\includegraphics[scale=.35]{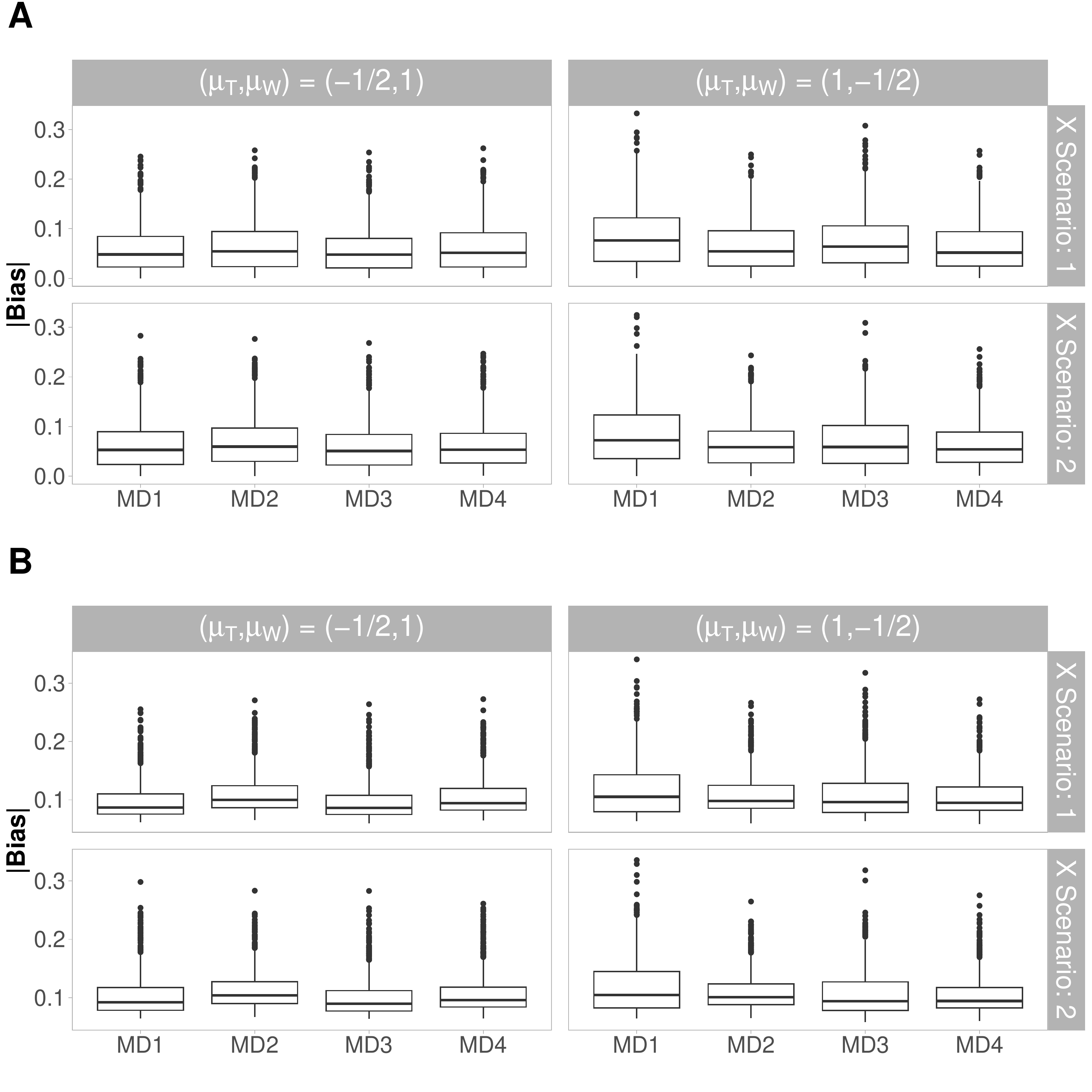}
	\caption{Boxplots of the absolute bias (Panel A) and RMSE (Panel B) of the ATE for the models described in Table \ref{tab:models_sim_study_binOut}. The results are based on 1000 Monte Carlo replicates. The columns correspond to variations of $\mu_T$ and $\mu_W$; rows correspond to the two scenarios of X distributions in the true data generation mechanism. We considered ${\rm Corr}(T_j,W_j)=0.5$.}
	\label{fig:ate2app}
\end{figure}

Figure \ref{fig:ate2app} displays boxplots of the absolute bias and RMSE of the ATE
for models listed in Table \ref{tab:models_sim_study_binOut}. The result support the idea that the ``best approach'' is sensitive to the data generating mechanism with $\mu_T$ and $\mu_W$ also playing a role in this discussion.
Specifically, MD1 and MD3 have lower bias values than MD2 and MD4 for $(\mu_T,\mu_W)^\top = (-1/2,1)^\top$, but the opposite occurs for $(\mu_T,\mu_W)^\top = (1,-1/2)^\top$. Additionally, models with random effects in the outcome model are less biased than their counterparts without random effects.

\section{TB Data Analysis}
	In this section we provide supplementary information about the analysis of the TB data.

	\subsection{Outcome Data Description}\label{app:outDef}
The data contains new cases of TB with a concluding diagnosis notified by the Brazilian public health system in 2016. In particular, we focus our analysis on individuals of at least 11 years of age.
In the full data frame, the column named \texttt{SITUA\_ENCE} gives the concluding diagnosis following the treatment for TB. A description this variable is given in Table \ref{tab:out}.

\begin{table}[!htp]
	\centering
	\caption{Description of the Outcome variable.}
	\begin{tabular}{c|l}
		\hline
		Variable Name & Values (Dictionary)\\
		\hline
		& 1 - Cure; 2 - Abandon\tablefootnote{\texttt{SITUA\_ENCE = 2} means that an individual abandon the treatment against TB after following it for at least 30 days.}; 3 - Death by TB; 4 - Death by other causes; \\
		\multirow{1}{*}{\texttt{SITUA\_ENCE}} &  5 - Transference; 6 - Change of diagnosis; 7 - Drug Resistance;\\
		&  8 - Change of scheme; 9 - Death during the treatment; 10 - Primary abandon\\
		\hline
	\end{tabular}
	\label{tab:out}
\end{table}

We focused our analyses on individuals who followed TB treatment for at least 30 days, which states that individuals recorded as primary abandon were excluded from the data analysis. We also excluded individuals who were lost to follow up due to death or moving out of the study region, as well as any patients who were marked as having a change of diagnosis. The data analysis only included individuals with \texttt{SITUA\_ENCE} recorded in $\{1,2,3,4,7,8\}$. The outcome was defined as $Y=1$ if \texttt{SITUA\_ENCE} is 1, and $Y=0$ if \texttt{SITUA\_ENCE} is 2 (Abandon), 3 (Death by TB), 4 (Death by other causes), 7 (Drug Resistance), or 8 (Change of scheme).

	\subsection{Exploratory Data Analysis}
	The TB data contain a total of $12,057$ observation in which $6,929$ individuals were assigned to DOT. Table \ref{tab:app_eda} presents a summary of the observed data for each of the variables available in the dataset. The row ``cured'' relates to the outcome variable, from which we observe that a total of $10,462$ individuals had a concluding diagnosis of cure at the end of the TB treatment, with $6,774$ corresponding to the number of individuals who were cured following DOT.
	\clearpage
	
	\begin{table}[!htb]
		\centering
		\caption{Baseline variables present in the analysis of the TB data.}
		\begin{tabular}{l|ccc}
			\hline
			\multirow{2}{*}{Variable} & Total & DOT & no DOT \\
			& ($12,057$ individuals)& ($6,929$ individuals) & ($5128$ individuals) \\
			\hline
			Cured, no.~(\%) & $10462\left( 87\% \right)$ & $6774\left( 98\%\right)$ & $3688\left( 72\%\right)$ \\
			AIDS, no.~(\%)       & $730\left( 6\% \right)$     & $289\left(4\%\right)$   & $441\left(9\%\right)$ \\
			Alcoholism, no.~(\%)  & $2103\left( 17\% \right)$    & $1136\left(16\%\right)$  & $967\left(19\%\right)$ \\
			Diabetes, no.~(\%)   & $777\left(6\% \right)$     & $423\left( 6\%\right)$   & $354\left(7\%\right)$ \\
			Drug Use, no.~(\%)   & $1969\left( 16\% \right)$    & $1069\left(15\%\right)$  & $900\left(17\%\right)$ \\
			Homelessness, no.~(\%)   & $346\left( 3\% \right)$     & $163\left(2\%\right)$   & $183\left(4\%\right)$ \\
			Male, no.~(\%)       & $8855\left(73\% \right)$    & $5156\left(74\%\right)$  & $3699\left( 72\%\right)$ \\
			Mental Illness, no.~(\%)    & $177\left(1\% \right)$     & $94\left(1\%\right)$    & $83\left( 2\%\right)$ \\
			In prison, no.~(\%)   & $2148\left(18\% \right)$    & $1381\left(20\%\right)$  & $767\left(15\%\right)$ \\
			Smoker, no.~(\%)     & $2966\left( 24\% \right)$    & $1687\left(24\%\right)$  & $1279\left(25\%\right)$ \\
			TB-ExtPulm, no.~(\%) & $1541\left(13\% \right)$    & $848\left(12\%\right)$   & $693\left(13\%\right)$ \\
			TB-Pulm, no.~(\%)    & $10229\left( 85\% \right)$   & $5956\left(86\%\right)$  & $4273\left(83\%\right)$  \\
			Age, mean~(SD)       & $37.5 \left(15.4\right)$  & $36.9 \left(15.1\right)$  & $38.3 \left(15.8\right)$\\
			HDI, mean~(SD)       & $0.779 \left(0.03\right)$ & $0.778 \left(0.03\right)$ & $0.781 \left(0.03\right)$\\
			\hline
		\end{tabular}
		\label{tab:app_eda}
	\end{table}
	In the dataset, the variable ``type of TB'' represents a categorical variable with three levels: pulmonary, extra-pulmonary and both (pulmonary + extra-pulmonary).  The data also have available indicator variables for diagnosis of each of AIDS, alcoholism, diabetes, or a mental illness, whether using illicit drugs, whether homeless, gender (male), whether currently in prison, and current smoking status. Further, the data also contain the age (years) of each individual. Finally, the human development index (HDI) of the city in which each individual lives is available.
	
	Next, we present an exploratory analysis of the balancing diagnostic for the observed covariates and the validity of the positivity assumption.

	\subsection*{Standardized Mean Difference}\label{app:smd}
	
	{\bf Definition 1. (SMD)} For a continuous variable $X$, let $\bar{X}_{treated}$ and $\bar{X}_{control}$ be the sample
	mean of $X$ in treated and control subjects, respectively. Moreover, let ${S}_{treated}^2$ and ${S}^2_{control}$ denote the sample
	variance of $X$ in treated and control subjects, respectively. The standardized mean difference (SMD) is defined as
	\begin{equation*}
		d = \frac{\bar{X}_{treated} - \bar{X}_{control}}{\sqrt{\frac{S^2_{treated} + S^2_{control}}{2}}}.
		\label{eq:smd_cont}
	\end{equation*}
	Although the SMD is developed for continuous variables, it can also be extended for dichotomous variables through the following expression \citep{austin2009using}
	\begin{equation*}
		d = \frac{\widehat{p}_{treated} - \widehat{p}_{control}}{\sqrt{\frac{\widehat{p}_{treated}(1-\widehat{p}_{treated}) + \widehat{p}_{control}(1-\widehat{p}_{control})}{2}}},
		\label{eq:smd_bina}
	\end{equation*}
	where $\widehat{p}_{treated}$ and $\widehat{p}_{control}$ are the sample prevalence of the variables in the treated and control groups.
	
	An alternative to the above quantities is to replace the sample means and sample variances by their weighted equivalents.
	The weighted mean is defined as $\bar{X}_{weight} = \frac{\sum_{i=1}^{n} \omega_iX_i}{\sum_{i=1}^{n} \omega_i}$, while the weighted variance is defined as $S^2_{weight} = \frac{\sum_{i=1}^{n} \omega_i}{\left(\sum_{i=1}^{n} \omega_i\right)^2 - \sum_{i=1}^{n} \omega_i^2}\sum_{i=1}^n\omega_i(X_i - \bar{X}_{weight})^2$, where $\omega_i$ is the weight of subject $i$.
	Usually, values of the absolute SMD lower than $10\%$ are taken to be an indication of reasonable balance \citep{austin2009balance}.
	
\begin{figure}[!htp]
		\begin{center}
			\includegraphics[width=6in]{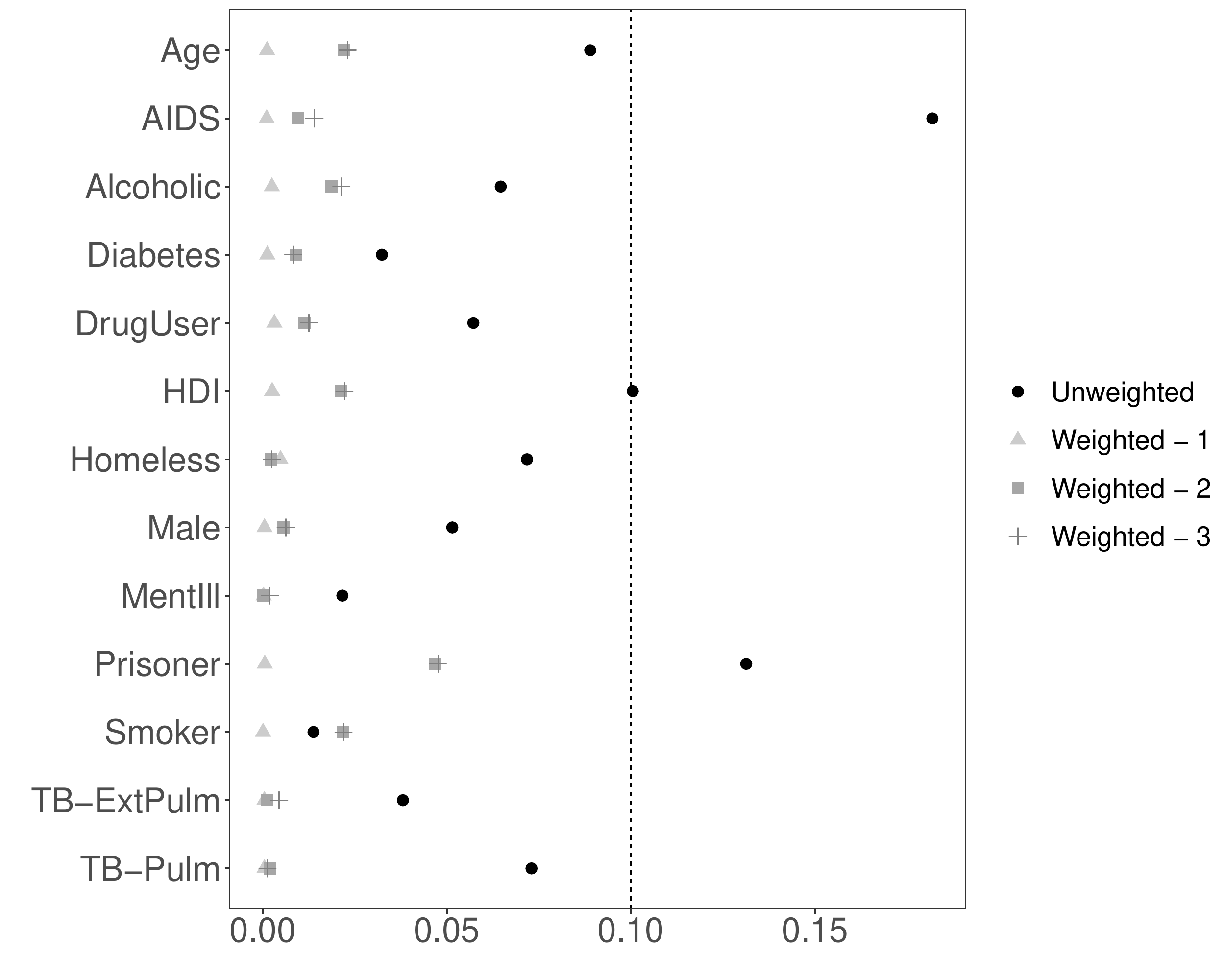}
		\end{center}
		\caption{Standardized mean difference (SMD) for the observed baseline covariates between treated and control subjects.}\label{fig:SMD-TB}
	\end{figure}

	Figure \ref{fig:SMD-TB} presents the absolute SMD evaluated for each covariate available in the data analysis. The weighted values were calculated assuming that $\omega =  Z/\widehat{PS} + (1-Z)/(1-\widehat{PS})$, where $Z$ is the observed exposure value for a particular unit and $\widehat{PS}$ corresponds to the estimated propensity score of the same particular unit. In the legend, Weighted - 1, Weighted - 2 and Weighted - 3 indicate the exposure model used to estimate the propensity score, namely PS1, PS2, and PS3.

	\clearpage
	
	\subsection*{Positivity Assumption}\label{app:positivity}
\begin{figure}[!htp]
		\centering
		\includegraphics[scale=.4]{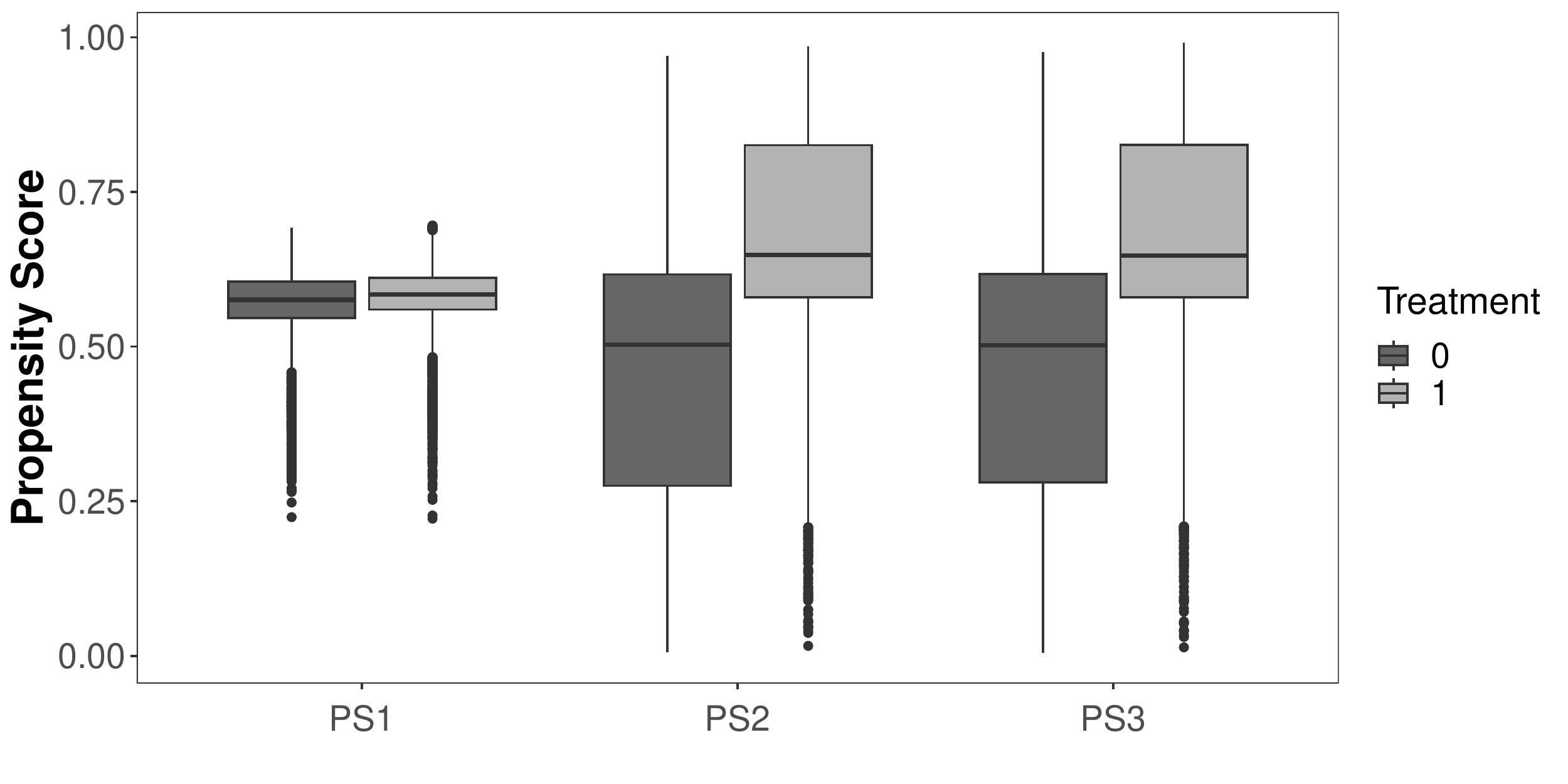}
		\caption{Propensity score distribution among treated and untreated units under three propensity score model specifications.}
		\label{fig:PS-overlap}
	\end{figure}

	Violation of the positivity assumption occurs when some group or strata are never (or rarely) assigned to some value of the treatment \citep{petersen2012diagnosing}. For instance, given a binary treatment $Z$ and a set of covariates $X$, the positivity assumption states that every sample unit has a positive probability of being assigned any value of the treatment, i.e.,~$P(Z={\rm z}|X) > 0$ for ${\rm z} =0,1$. For propensity score methods, this assumption is key because the methods aim, broadly, at comparing individuals with the same value of the propensity score but different values of the treatment assignment. So, if there is a stratum in the sample that never (or always) receives a particular treatment value, causal effects might be non-identifiable. Figure \ref{fig:PS-overlap} shows boxplots of the propensity score estimated for each of the proposed exposure models and separated according to the assignment of DOT. Although the propensity distributions among treated and untreated units are more similar in model PS1 than PS2 and PS3, all models provide an acceptable overlap.
	
	\clearpage

	\subsection{Model Fit Criteria}\label{app:tb-modelcomp}
	Here we present some model comparison criteria for the exposure (Table \ref{tab:waic_looEXP}) and outcome (Table \ref{tab:waic_looOUT}) models.
	
	\begin{table}[!htp]
		\centering
		\caption{Exposure model comparison.}
		\begin{tabular}{l|ccc|ccc}
			\hline
			& elpd\tablefootnote{elpd: expected log predictive density}~(WAIC\tablefootnote{WAIC: Watanabe-Akaike information criterion}) & pWAIC\tablefootnote{pWAIC (and pLOO): estimated value of the effective number of parameters} & WAIC & elpd~(LOO\tablefootnote{LOO: leave-one-out}) & pLOO${}^{3}$ & LOO \\
			\hline
			PS1 & -8136.17 & 14.10 & 16272.35 & -8136.23 & 14.16 & 16272.45 \\
			PS2 & -6884.15 & 242.63 & 13768.30 & -6893.51 & 251.99 & 13787.02 \\
			PS3 & -6877.74 & 235.82 & {\bf 13755.49} & -6885.21 & 243.28 & {\bf 13770.42} \\
			\hline
		\end{tabular}
		\label{tab:waic_looEXP}
	\end{table}
	Metrics of model comparison show how well the model fits the data. Therefore, for causal consideration, this analysis may not be appropriate when discussing the balancing induced for confounders - in such a context, the SMDs are more appropriate for assessing the fit of the propensity score models. The target of this analysis is to understand the strength of the structure recovered by the cluster-level random effect in the propensity score and outcome models.
	
	\begin{table}[!htp]
		\centering
		\caption{Outcome model comparison.}
		\begin{tabular}{l|ccc|ccc}
			\hline
			& elpd${}^{1}$~(WAIC${}^{2}$) & pWAIC${}^{3}$ & WAIC & elpd~(LOO${}^{4}$) & pLOO${}^{3}$ & LOO \\
			\hline
			M1 & -3788.84 & 1.99 & 7577.67 & -3788.84 & 2.00 & 7577.69 \\
			M2 & -3622.66 & 120.45 & 7245.32 & -3623.69 & 121.48 & 7247.38 \\
			M3 & -3622.60 & 116.19 & 7245.20 & -3623.55 & 117.15 & 7247.10 \\
			M4 & -3538.21 & 15.07 & 7076.42 & -3538.27 & 15.13 & 7076.54 \\
			M5 & -3435.76 & 115.90 & {\bf 6871.52} & -3436.61 & 116.74 & {\bf 6873.21} \\
			M6 & -3436.19 & 111.72 & 6872.37 & -3436.94 & 112.47 & 6873.88 \\
			M7 & -3649.03 & 2.88 & 7298.06 & -3649.04 & 2.89 & 7298.07 \\
			M8 & -3535.18 & 104.18 & 7070.36 & -3535.90 & 104.90 & 7071.81 \\
			M9 & -3535.41 & 101.62 & 7070.82 & -3536.10 & 102.31 & 7072.19 \\
			M10 & -3734.95 & 2.90 & 7469.90 & -3734.96 & 2.91 & 7469.92 \\
			M11 & -3612.22 & 138.22 & 7224.45 & -3613.87 & 139.87 & 7227.74 \\
			M12 & -3613.57 & 131.75 & 7227.15 & -3615.02 & 133.20 & 7230.05 \\
			M13 & -3735.10 & 2.99 & 7470.21 & -3735.11 & 3.00 & 7470.23 \\
			M14 & -3613.55 & 137.26 & 7227.09 & -3615.14 & 138.86 & 7230.28 \\
			M15 & -3614.23 & 131.34 & 7228.47 & -3615.66 & 132.77 & 7231.32 \\
			\hline
		\end{tabular}
		\label{tab:waic_looOUT}
	\end{table}

	\clearpage

	\subsection{Sensitivity Analysis}
	The original dataset contains missing values for four covariates: ethnicity, schooling, immigration status, and healthcare profession. Due to the high number of missing values, these covariates were excluded from the data analysis presented in the main text. A sensitive analysis was performed by comparing the data analysis presented in the main text with the one that excludes rows with missing values on the covariates ethnicity, schooling, immigration status, and healthcare profession. The final dataset, after excluding rows with missing values on such covariates, contains 7951 individuals across 373 cities.

	Tables \ref{tab:sens1} and \ref{tab:sens2} compare the SMD values and estimates of ATE and odds ratio (OR) between Analysis 1 (excluding rows with missing data on covariates) and Analysis 2 (presented in the main text).
	First, the unweighted SMD values from those covariates excluded in Analysis 2 are all lower than 5\% in Analysis 1.
	Further, in general, the remaining values agree in magnitude.
	Regarding estimates of the ATE and OR, the absolute relative difference between the models that include a random effect in the outcome model and its respective counterpart that does not is also shown. The results suggest minimal differences between both analyses, since the ATE, OR, and absolute relative difference provide similar values.
	\begin{table}[!htb]
		\centering
		\caption{Standardized mean difference (SMD) for the observed covariates between treated and control units. Analysis 1 excludes, from the analytic dataset, rows that have missing values on covariates ethnicity, schooling, if immigrant, and if a healthcare professional. Analysis 2 excludes covariates ethnicity, schooling, if immigrant, and if a healthcare professional from the data analysis. Analysis 1 contains 7,951 individuals across 373 cities, whereas Analysis 2 includes 12,057 individuals spread over 415 cities.}
		\begin{tabular}{l|cccc|cccc}
			\hline
			&\multicolumn{4}{c}{Analysis 1} & \multicolumn{4}{c}{Analysis 2}\\
			Variables& Unw. & Weig. - 1 & Weig. - 2 & Weig. - 3& Unw. & Weig. - 1 & Weig. - 2 & Weig. - 3 \\
			\hline
			Male & 0.03 & 0.00 & 0.01 & 0.01 & 0.05 & 0.00 & 0.01 & 0.01\\
			TB-Pulm & 0.07 & 0.00 & 0.01 & 0.00  & 0.07 & 0.00 & 0.00 & 0.00\\
			TB-ExtPulm & 0.03 & 0.00 & 0.00 & 0.00  & 0.04 & 0.00 & 0.00 & 0.00\\
			AIDS & 0.15 & 0.00 & 0.01 & 0.01  & 0.18 & 0.00 & 0.01 & 0.01\\
			Alcoholism & 0.09 & 0.00 & 0.03 & 0.03  & 0.06 & 0.00 & 0.02 & 0.02\\
			Diabetes & 0.05 & 0.00 & 0.00 & 0.00  & 0.03 & 0.00 & 0.01 & 0.01\\
			Mental Illness & 0.02 & 0.00 & 0.01 & 0.01 & 0.02 & 0.00 & 0.00 & 0.00 \\
			Illicit drug use & 0.06 & 0.00 & 0.01 & 0.01  & 0.06 & 0.00 & 0.01 & 0.01 \\
			Smoker & 0.03 & 0.00 & 0.04 & 0.03 & 0.01 & 0.00 & 0.02 & 0.02\\
			In prison & 0.14 & 0.00 & 0.05 & 0.05 & 0.13 & 0.00 & 0.05 & 0.05 \\
			Homelessness & 0.10 & 0.01 & 0.00 & 0.00  & 0.07 & 0.00 & 0.00 & 0.00 \\
			Age & 0.05 & 0.00 & 0.03 & 0.02  & 0.09 & 0.00 & 0.02 & 0.02\\
			HDI & 0.14 & 0.00 & 0.04 & 0.04 & 0.10 & 0.00 & 0.02 & 0.02\\
			Health Worker & 0.01 & 0.00 & 0.01 & 0.01 & --& --& --& -- \\
			Immigrant & 0.04 & 0.00 & 0.00 & 0.00 & --& --& --& --\\
			Ethnicity A & 0.01 & 0.00 & 0.00 & 0.01 & --& --& --& --\\
			Ethnicity B & 0.02 & 0.00 & 0.00 & 0.00 & --& --& --& --\\
			Ethnicity C & 0.05 & 0.00 & 0.01 & 0.01 & --& --& --& --\\
			Ethnicity D & 0.03 & 0.00 & 0.00 & 0.00 & --& --& --& --\\
			Sch. level 1 & 0.00 & 0.00 & 0.01 & 0.01 & --& --& --& --\\
			Sch. level 2 & 0.01 & 0.00 & 0.01 & 0.00 & --& --& --& --\\
			Sch. level 3 & 0.03 & 0.00 & 0.01 & 0.01 & --& --& --& --\\
			Sch. level 4 & 0.00 & 0.00 & 0.02 & 0.02 & --& --& --& --\\
			\hline
		\end{tabular}
		\label{tab:sens1}
	\end{table}

	\begin{table}[!htb]
		\centering
		\caption{Estimates of the ATE and OR for the models described in Table \ref{tab:modelsTB}. Analysis 1 excludes, from the analytic dataset, rows that have missing values on covariates ethnicity, schooling, if immigrant, and if a healthcare professional. Analysis 2 excludes covariates ethnicity, schooling, if immigrant, and if a healthcare professional from the data analysis. Columns Relative ATE and Relative OR correspond to the absolute relative difference between models that include a random effect in the outcome model and its respective counterparts that do not. Analysis 1 contains 7,951 individuals across 373 cities, whereas Analysis 2 includes 12,057 individuals spread over 415 cities.}
		\begin{tabular}{l|cccc|cccc}
			\hline
			& \multicolumn{4}{c}{ATE}& \multicolumn{4}{c}{OR} \\
			& \multicolumn{2}{c}{Analysis 1} & \multicolumn{2}{c}{Analysis 2} & \multicolumn{2}{c}{Analysis 1} & \multicolumn{2}{c}{Analysis 2}  \\
			& ATE & Relative ATE  & ATE & Relative ATE & OR & Relative OR  & OR & Relative OR \\
			\hline
			M1 & 0.27 & --    & 0.26 & --   & 17.96 & --   & 17.16 &  -- \\
			M2 & 0.30 & 0.11  & 0.29 & 0.13 & 25.62 & 0.43 & 25.53 & 0.49 \\
			M3 & 0.30 & 0.11  & 0.29 & 0.13 & 25.52 & 0.42 & 25.29 & 0.47 \\
			M4 & 0.26 & --    & 0.24 & --   & 17.97 & --   & 17.52 & -- \\
			M5 & 0.28 & 0.11  & 0.27 & 0.12 & 24.98 & 0.39 & 24.58 & 0.40\\
			M6 & 0.28 & 0.11  & 0.27 & 0.12 & 25.02 & 0.39 & 24.19 & 0.38\\
			M7 & 0.26 & --    & 0.24 & --   & 16.75 &  --  & 16.10 & -- \\
			M8 & 0.29 & 0.11  & 0.27 & 0.12 & 23.39 & 0.40 & 22.60 & 0.40\\
			M9 & 0.29 & 0.11  & 0.27 & 0.12 & 23.25 & 0.39 & 22.55 & 0.40\\
			M10 & 0.32 & --   & 0.30 & --   & 25.24 & --   & 23.64 & --\\
			M11 & 0.30 & 0.04 & 0.29 & 0.03 & 26.06 & 0.03 & 24.87 & 0.05 \\
			M12 & 0.30 & 0.04 & 0.29 & 0.03 & 25.84 & 0.02 & 24.91 & 0.05\\
			M13 & 0.32 & --   & 0.30 & --   & 25.20 &  --  & 23.65 & --\\
			M14 & 0.30 & 0.04 & 0.29 & 0.03 & 26.02 & 0.03 & 25.15 & 0.06\\
			M15 & 0.30 & 0.04 & 0.29 & 0.03 & 25.92 & 0.03 & 24.91 & 0.05\\
			\hline
		\end{tabular}
		\label{tab:sens2}
	\end{table}
	\clearpage

	\subsection{Comparison with a Joint Approach}
	As pointed out in the main text, there have been several proposals of Bayesian procedures to estimate causal effects based on propensity score methods. In particular, \cite{Stephens2022} addresses this point, and demonstrates that a full Bayesian solution can be obtained following a plug-in, two-step approach: we adopt this proposal in the paper.
	
	Here, we illustrate the performance of a joint approach in the data analysis. In particular, we considered the following variation of the joint model:
	\begin{equation}
		\begin{split}
			{\rm logit}(P(Y_{ji} = 1|Z_{ji},\mathbf{X}_{ji})) & = \beta + Z_{ji}\beta_Z + \mathbf{X}^\top_{ji}\bfbeta_b + \eta_j\\
			{\rm logit}(P(Z_{ji} = 1|\mathbf{X}_{ji}))  &= \sum_{k=1}^{q} \gamma_kX_{kji} + \eta_j,
		\end{split}
	\end{equation}
	where $\eta_j$ is a shared random effect. We considered two specifications for $\eta_j$:
	\begin{itemize}
		\item[] M16: $\bfeta| \phi \sim \mathcal{N}(\mathbf{0},\phi^2\mathbf{I}_N)$,
		\item[] M17: $\bfeta | \phi,\vartheta \sim \mathcal{N}(\mathbf{0},\phi^2\mathbf{R}(\vartheta))$, where $R_{lk} = {\rm Corr}(\eta_{l},\eta_{k}) = {\rm exp}\left(-\vartheta ||\bfs_l - \bfs_k|| \right)$, with $\bfs_k$ denoting the centroid of city $k$ and $||\cdot||$ denoting the Euclidean distance.
	\end{itemize}
	
	A summary of the ATE posterior distribution of such models were compared with models M5 and M6 described in Table \ref{tab:modelsTB}. The results are summarized in Table \ref{tab:rev_joint}. Clearly, the results do not agree between models M5 and M16, and between M6 and M17. While the true values are not known in the real data analysis, the work of \cite{Stephens2022} suggests that the joint modelling approach is incorrect and leads to bias, and the differences in results here could reasonably be attributed to the use of an inappropriate (joint) model.
	
	\begin{table}[!htb]
		\centering
		\caption{Summary of the ATE posterior distributions for models M5, M6, M16, and M17.}
		\begin{tabular}{l|cccc}
			\hline
			Models &  Mean  &  SD &quant. 2.5\% &quant. 97.5\%\\
			\hline
			M5 & 0.272& 0.007    &   0.258 &       0.287\\
			M6  &0.271& 0.007  &     0.259   &     0.285\\
			M16 &0.198& 0.006   &    0.186  &      0.210\\
			M17 &0.198& 0.006 &      0.187    &    0.209\\
			\hline
		\end{tabular}
		\label{tab:rev_joint}
	\end{table}
	
\begin{figure}[!htb]
		\centering
		\includegraphics[scale=.34]{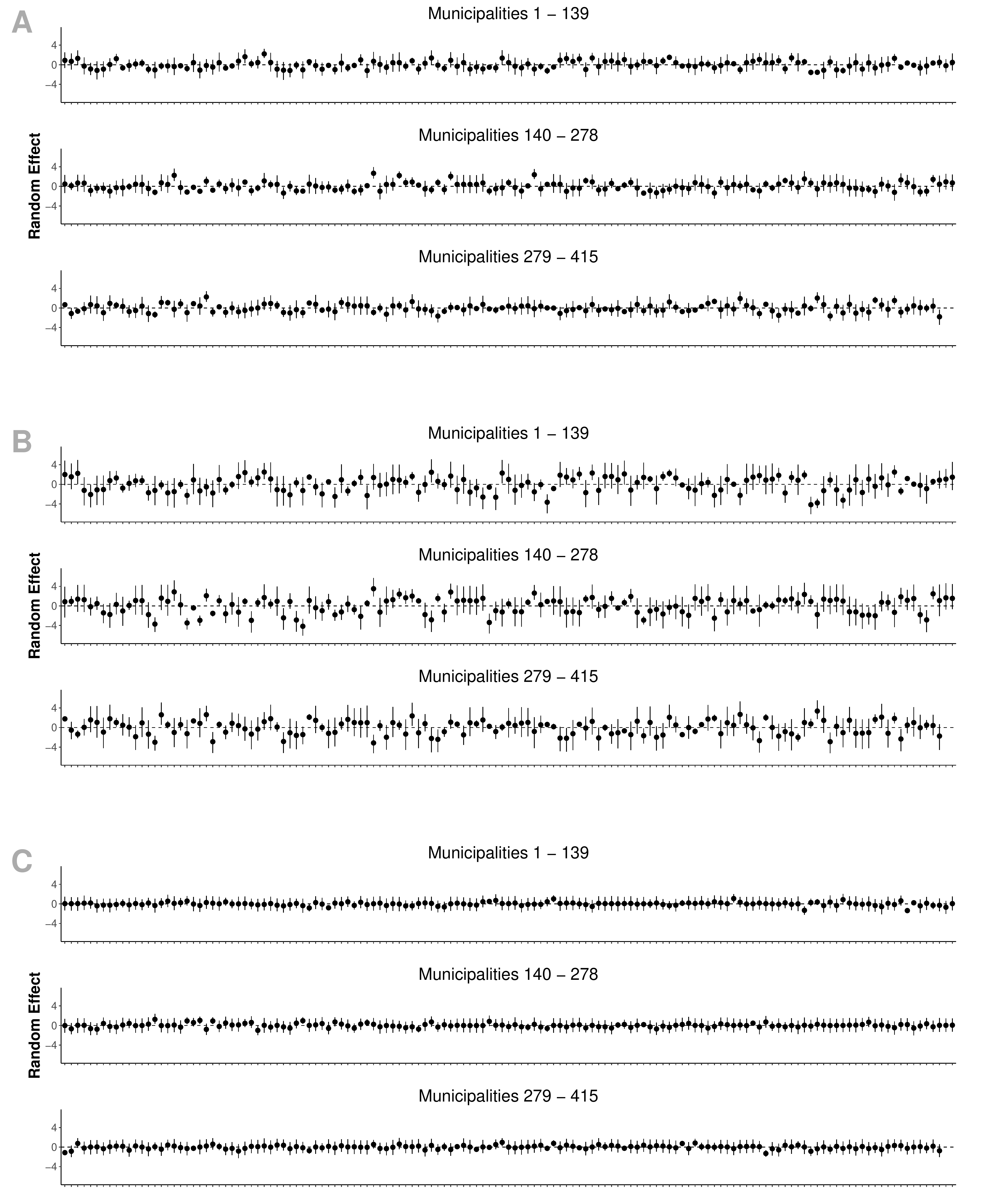}
		\caption{Summary of the random effects posterior distribution for the 415 municipalities estimated from models M16 (Panel A), PS2 (Panel B) and M5 (Panel C). The vertical lines are the 95\% credible intervals, whereas the circles represent the posterior mean. The horizontal dashed line is fixed at zero.}
		\label{fig:rand_eff1}
	\end{figure}

\begin{figure}[!htb]
		\centering
		\includegraphics[scale=.34]{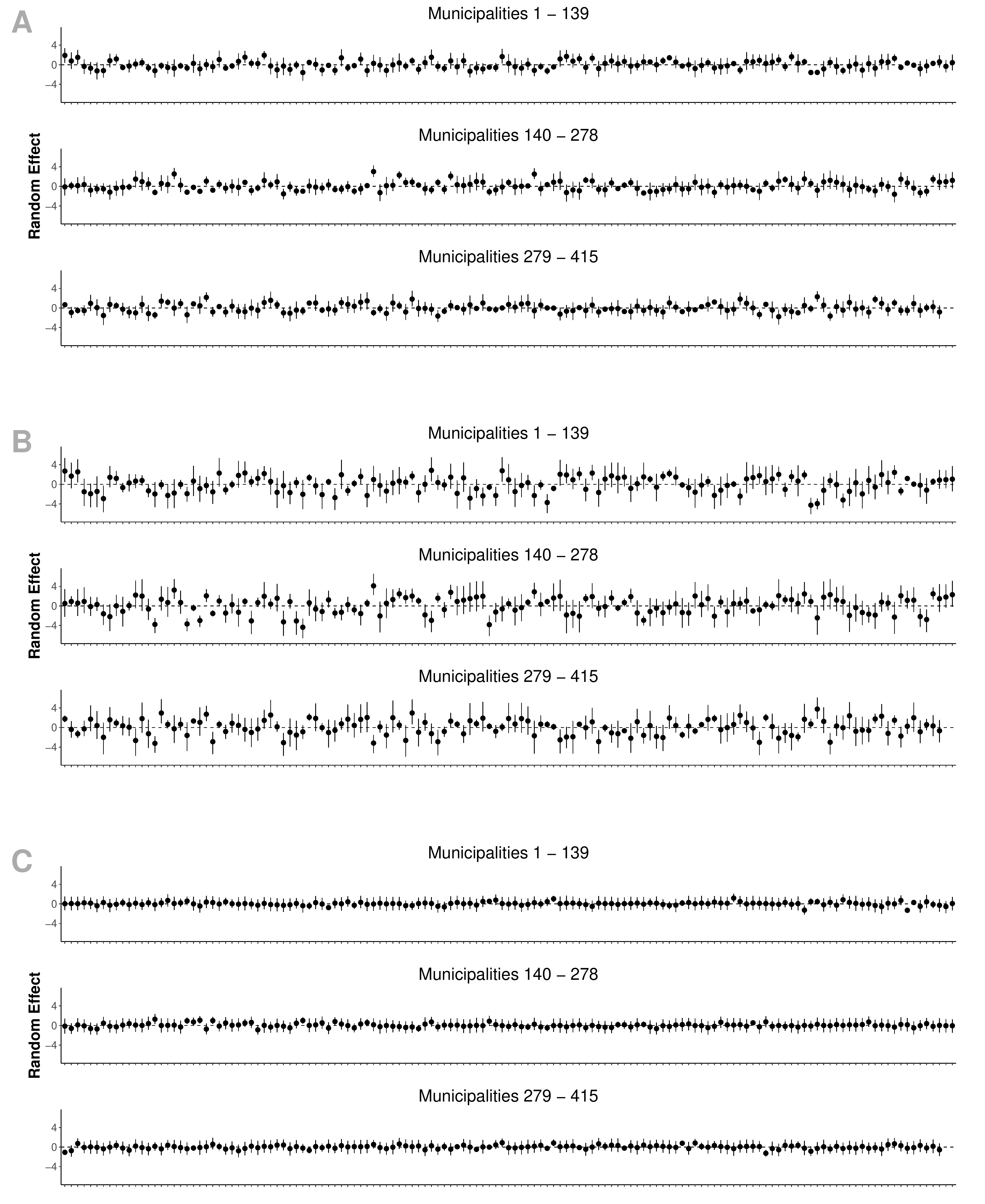}
		\caption{Summary of the random effects posterior distribution for the 415 municipalities estimated from models M17 (Panel A), PS3 (Panel B) and M6 (Panel C). The vertical lines are the 95\% credible intervals, whereas the circles represent the posterior mean. The horizontal dashed line is fixed at zero.}
		\label{fig:rand_eff2}
	\end{figure}

	Figures \ref{fig:rand_eff1} and \ref{fig:rand_eff2} compare a summary of the random effects posterior distribution for the 415 municipalities estimated from models M5 and M16 (Figure \ref{fig:rand_eff1}), and M6 and M17 (Figure \ref{fig:rand_eff2}). Vertical lines are the 95\% credible interval and points are the posterior mean for each random effect. We observe several effects with credible intervals totally below or above the zero horizontal dashed line. In both figures, Panel A displays the summary of the random effects estimated from the joint approaches M16 (Figure \ref{fig:rand_eff1}) and M17 (Figure \ref{fig:rand_eff2}); Panel B shows the summary of the random effects estimated from the exposure model PS2 (Figure \ref{fig:rand_eff1}) and PS3 (Figure \ref{fig:rand_eff2}); Panel C displays the summary of the random effects estimated from the outcome model M5 (Figure \ref{fig:rand_eff1}) and M6 (Figure \ref{fig:rand_eff2}).
	Clearly, the joint method provides different summaries from those obtained in the two-step procedure.
	
	\clearpage
	
	\subsection{Additional information for models PS3 and M6}
	
	\begin{figure}[!htb]
		\centering
		\includegraphics[scale=.35]{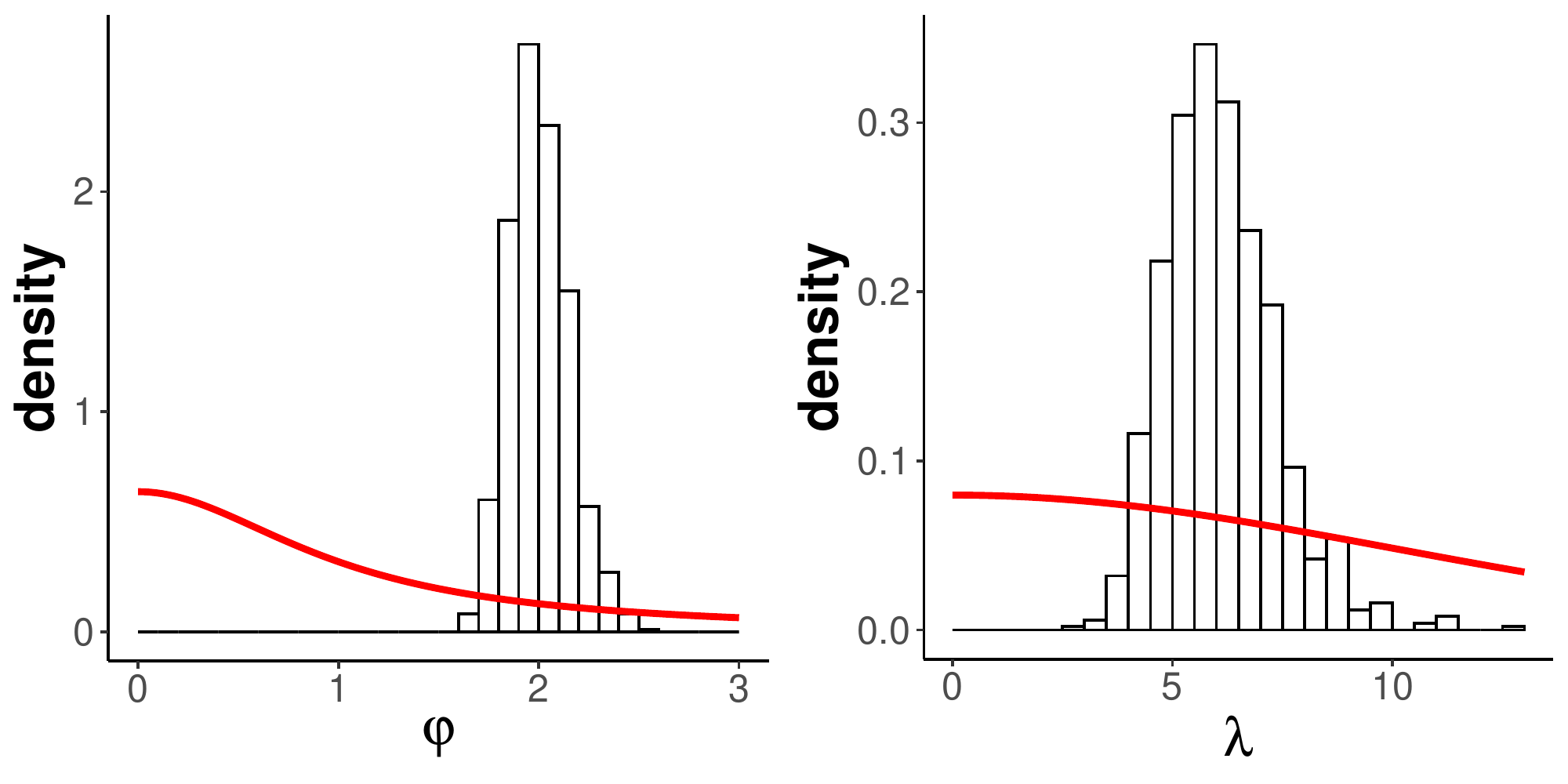}
		\caption{Histograms of the hyperparameters presented in model PS3. The red lines are the prior densities.}
		\label{fig:hpPS3}
	\end{figure}
	
	\begin{figure}[!htb]
		\centering
		\includegraphics[scale=.35]{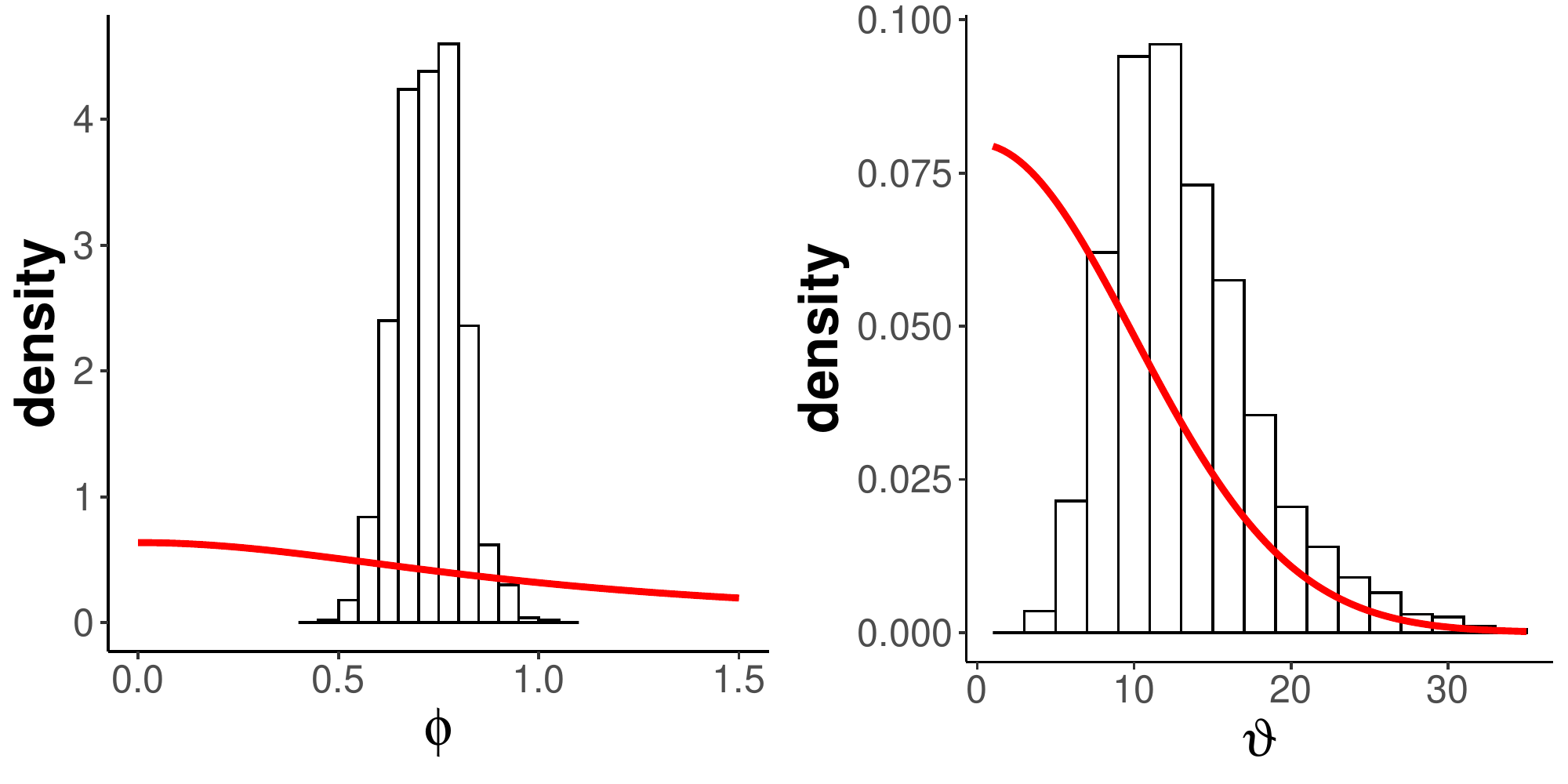}
		\caption{Histograms of the hyperparameters presented in model M6. The red lines are the prior densities.}
		\label{fig:hpM6}
	\end{figure}
	
\end{document}